\documentclass[%
 reprint,
nofootinbib,
 amsmath,amssymb,
 aps,
]{revtex4-1}

\usepackage{soul}
\usepackage{graphicx}
\usepackage[dvipsnames]{xcolor}
\usepackage{dcolumn}
\usepackage{bm}
\usepackage[colorlinks]{hyperref}
\usepackage{mathrsfs}
\usepackage{verbatim}
\usepackage{float}
\usepackage{subcaption} 
\usepackage{amsfonts, pifont}
\usepackage[inline]{enumitem}
\newlist{inlinelist}{enumerate*}{1}
\setlist*[inlinelist,1]{%
  label=(\arabic*),
}
\usepackage[normalem]{ulem}

\usepackage{afterpage}

\renewcommand{\arraystretch}{1.25}
\newcommand{\bear}{\begin{array}}
\newcommand{\ear}{\end{array}}

\newcommand{\beq}{\begin{eqnarray}}
\newcommand{\eeq}{\end{eqnarray}}
\newcommand{\beqa}{\begin{eqnarray}}
\newcommand{\eeqa}{\end{eqnarray}}

\newcommand{\nn}{\nonumber}
\def\OMIT#1{{}}
\newcommand{\lsim}{\mathrel{\rlap{\lower4pt\hbox{\hskip1pt$\sim$}}
    \raise1pt\hbox{$<$}}}         
\newcommand{\gsim}{\mathrel{\rlap{\lower4pt\hbox{\hskip1pt$\sim$}}
    \raise1pt\hbox{$>$}}}         

\newcommand{\Ol}{{\cal O}}

\newcommand{\bl}{\left}
\newcommand{\br}{\right}

\usepackage{mathtools}

\newcommand{\GeV}{{\rm GeV}}
\newcommand{\MeV}{{\rm MeV}}

\newcommand{\eV}{{\rm eV}}

\newcommand{\kg}{{\rm kg}}
\newcommand{\cm}{{\rm cm}}
\newcommand{\km}{{\rm km}}
\newcommand{\seg}{{\rm s}}
\newcommand{\yr}{{\rm year}}

\newcommand{\DM}{{\rm DM}}
\newcommand{\crs}{{\rm crystal}}

\newcommand{\vmin}{v_{\rm min}}
\newcommand{\vesc}{v_{\rm esc}}
\newcommand{\vmp}{v_{\rm mp}}

\newcommand{\cov}{{\bm\Sigma}}

\newcommand{\dd}{\mathrm{d}}
\newcommand{\Sun}{\odot}
\newcommand{\Earth}{\oplus}

\usepackage{diagbox}
\usepackage[flushleft]{threeparttable}
\usepackage{multirow}
\newcommand{\loge}{\text{ln}}
\newcommand{\sig}{\mathbf{S}}
\newcommand{\bkg}{\mathbf{B}}

\makeatletter
\newcommand*{\rom}[1]{\expandafter\@slowromancap\romannumeral #1@}
\usepackage{pbox}

\def\nn{\nonumber}

\def\Eq#1{Eq.~(\ref{#1})}
\def\Eqs#1{Eqs.~(\ref{#1})}

\newcommand\myshade{100}
\colorlet{mycitecolor}{blue}
\colorlet{myurlcolor}{violet}

\hypersetup{
  citecolor  = mycitecolor!\myshade!black,
  urlcolor   = myurlcolor!\myshade!black,
  colorlinks = true
}
\begin{document}

\title{Dark Matter Substructure under the Electron Scattering Lamppost}
\author{Jatan Buch} \email{jatan\textunderscore buch@brown.edu}
\author{Manuel A. Buen-Abad} \email{manuel\textunderscore buen-abad@brown.edu}
\author{JiJi Fan} \email{jiji\textunderscore fan@brown.edu}
\author{John Shing Chau Leung} \email{shing\textunderscore chau\textunderscore leung@alumni.brown.edu}
\affiliation{Department of Physics, Brown University, Providence, RI, 02912, USA}

\date{\today}

\begin{abstract}
We study the mutual relationship between dark matter-electron scattering experiments and possible new dark matter substructure nearby hinted by the {\it Gaia} data. We show how kinematic substructure could affect the average and modulation spectra of dark matter-electron scattering in semiconductors, and the discovery reaches of future experiments with these targets. Conversely, we demonstrate how future data could probe and constrain the substructure dark matter fraction, even when it constitutes a sub-dominant component of the local dark matter density.
\end{abstract}

\maketitle


%

\section{Introduction}
\label{sec:intro}

The search for dark matter (DM) is one of the main goals for both experimental and theoretical physics. Among different search strategies, direct detection experiments are some of the most ambitious. In recent years various collaborations have made significant progress in constraining the parameter space for DM interactions with the particles of the standard model (SM). Traditional nuclear recoil experiments \cite{Akerib:2016vxi, Agnese:2017njq, Cui:2017nnn, Aprile:2018dbl, Abdelhameed:2019hmk, Armengaud:2019kfj} are pushing the limits on the DM interaction cross section towards the neutrino floor for heavy DM particles, whereas electron recoil experiments \cite{Tiffenberg:2017aac, Crisler:2018gci, Abramoff:2019dfb, Arnaud:2020svb, Barak:2020fql, Aguilar-Arevalo:2019wdi, Aprile:2019xxb, Amaral:2020ryn} are probing increasingly smaller DM masses. For a comparison of different targets for sub-GeV DM direct detection, see~\cite{Griffin:2019mvc}.

It is well known that an accurate understanding of direct detection results depends crucially on the characteristics of the local DM distribution. Indeed, besides the local number density of DM particles, kinematic quantities such as DM mean velocity and its velocity dispersion are critical to the hope for a discovery at direct detection experiments.


The standard halo model (SHM), in which the velocities of the DM particles follow an isotropic Maxwellian distribution, has commonly been used in the computation of DM direct detection rates. Recent work, however, has shown that the Milky Way's history has been punctuated by mergers with dwarf galaxies, which resulted in a rich variety of {\it stellar} substructure beyond the traditional halo and disk, such as the debris flow of the {\it Gaia} Sausage (also called Enceladus) \cite{Necib:2018iwb, Necib:2018igl}, the Nyx stream \cite{Necib:2019zbk, Necib:2019zka}, or the so called ``shards'', the S1, S2a, and S2b streams \cite{10.1093/mnras/sty1403, OHare:2018trr, OHare:2019qxc}.\footnote{We caution the reader that the analysis extracting the {\it Gaia} Sausage substructure was performed in a region of the sky (within galactocentric radii of $7.5$--$10$ kpc and $\vert z \vert > 2.5$ kpc), slightly different from the analysis for the Nyx stream (within radii $6.5$--$9.5$ kpc and $\vert z \vert < 2$ kpc).} Since dwarf galaxies also contain DM, these mergers could also result in associated {\it dark matter} substructure within our galaxy, beyond that of what pertains to the SHM.

The impact that these new astrophysical discoveries could have on DM direct detection searches has only started to be explored recently~\cite{Savage:2006qr, OHare:2018trr, Wu:2019nhd, Buckley:2019skk, OHare:2019qxc, Buch:2019aiw}. In this paper, we focus on the effects that DM substructure hinted by these discoveries could have on electron recoil experiments with semiconductor targets. In particular, we are interested in how the differential DM-electron scattering rate depends on DM velocity distributions beyond the SHM,\footnote{For earlier studies on DM-electron scattering, see \cite{Kopp:2009et, Dedes:2009bk, Graham:2012su, Essig:2011nj, Lee:2015qva, Essig:2015cda, Andersson:2020uwc}.} as well as in the hitherto unexplored possibility of using the differential recoil rate to deduce the astrophysical properties of possible DM substructure components (see Ref.~\cite{Kavanagh:2020cvn} for an analysis similar in spirit to ours but with a different objective). A particularly interesting feature of direct detection searches that DM substructure can affect is the seasonal variation of a DM signal~\cite{Drukier:1986tm}. The presence of DM substructure could produce an annual modulation signatures differing in both amplitude and phase from what is expected for DM in the SHM \cite{Freese:2012xd, Froborg:2020tdh}.

In order to achieve these goals, we need to define the DM velocity distributions we will be using in our analysis. Besides the SHM, we will also consider the halo and Sausage distributions as inferred in Refs.~\cite{Necib:2018iwb, Necib:2018igl}. Indeed, there seems to be good evidence that low-metallicity stars originating from older mergers, such as those in the {\it Gaia} Sausage, are good kinematic tracers of DM and thus allow for their associated DM distribution to be determined up to uncertainties in the subtructure fraction~\cite{Necib:2018igl}.\footnote{This claim is contingent on the merger history of Milky Way-like galaxies in simulations~\cite{Bozorgnia:2018pfa, Lisanti:2018rcw, Bozorgnia:2019mjk}.} However, the correlation between the stellar streams, which arise from more recent mergers, and their associated DM is currently the object of some debate, and is far from being completely understood. Lacking an accurate description of how DM associated with the stellar streams is distributed, we will restrict ourselves to taking the velocity distributions of the stellar streams as benchmarks for their dark matter counterparts. To this end, we use the distributions for the Nyx~\cite{Necib:2019zbk}, S1, S2a, and S2b~\cite{OHare:2019qxc} stellar streams. Since we are interested in exploring the relationship between astrophysics and direct detection, we take the stellar streams as mere proxies or placeholders for their associated DM distributions and make no claims about their accuracy as such.


The paper is organized as follows. We revisit the formalism of the DM-electron scattering rate for semiconductor targets, describe the astrophysical setup we consider, and develop an intuition for the impact of various DM substructure components in Sec.~\ref{sec:dm_er}. Then we will present our statistical analysis in Sec.~\ref{sec:statistics}, and discuss our results on the discovery reaches assuming different DM velocity distributions as well as how future DM-electron experiments could probe fractions of substructure components in Sec.~\ref{sec:results}. We state our conclusions and the outlook of our work in Sec.~\ref{sec:discussion}. We include further details in four appendices.

\section{Dark Matter Electron Recoil Rate}
\label{sec:dm_er}

In this section, we first review the basic formalism for dark matter-electron (DM-$e$) scattering in semiconductors, which could be skipped by readers who are familiar with the subject. We then present a novel description of the effects that the DM velocity distributions from various substructures have on this type of scattering.

\subsection{Formalism}
\label{subsec:form}
The differential scattering rate of DM particles $\chi$ off the electrons in a semiconducting target material, or {\it spectrum} for brevity, is given by \cite{Graham:2012su, Essig:2011nj, Lee:2015qva, Essig:2015cda}:
\beq\label{diff_rate}
	\frac{\dd R}{\dd \ln E} = N_{\rm cell} \frac{\rho_\chi}{m_\chi} \overline{\sigma}_e \alpha ~ \kappa(E, t) \ ,
\eeq
where $R$ is the event rate per unit mass; $E$ is the total energy transferred to the electron; $N_{\rm cell}$ is the number of cells per unit mass of target material; $\rho_\chi \approx 0.4 ~ \GeV/\cm^3$ and $m_\chi$ are the local DM energy density~\cite{Sivertsson:2017rkp, Buch:2018qdr} and the DM mass respectively; $\overline{\sigma}_e$ parameterizes the DM-$e$ coupling;\footnote{$\overline{\sigma}_e$ corresponds exactly to the free elastic scattering cross section in the heavy meadiator case.} $t$ is the time of the year, and $\alpha$ is the QED coupling constant. The $\kappa(E, t)$ factor is a ``correction'' factor that takes into account the particular properties of the semiconducting target, the local DM velocity distribution, and the momentum dependence of the DM-$e$ interactions. It is given by \cite{Graham:2012su, Essig:2011nj, Lee:2015qva, Essig:2015cda}:
\beq\label{kappa_correction}
\begin{split}
	\kappa(E, t) = \frac{m_e^2}{\mu_{\chi e}^2} \int \dd q ~ \frac{E}{q^2} F_\DM^2(q) & \vert f_\crs(q,E) \vert^2 \\ & \times g(\vmin(q,E), t) \ ,
\end{split}
\eeq
with $\mu_{\chi, e}$ the DM-$e$ reduced mass; $q$ the momentum transfer, $F_\DM(q) \equiv \bl( \frac{\alpha m_e}{q} \br)^n$ the {\it DM form factor}, which parameterizes the momentum dependence of the DM-$e$ scattering; $\vert f_\crs(q,E) \vert^2$ the {\it crystal form factor}, which describes the response of the semiconductor material to be probed with momentum $q$ and energy $E$; and $g(\vmin, t)$ the mean inverse speed of those DM particles with speeds above the minimum $\vmin$ required to scatter off the target.\footnote{In this paper, we consider the simplest possibility that the DM-$e$ scattering is velocity independent. Otherwise, the definition of $g(\vmin, t)$ needs to be modified, along with the corresponding form factors following the treatment in Ref.~\cite{Catena:2019gfa}.} The time dependence arises from the annual modulation of the DM wind in the lab frame, due to the Earth's motion around the Sun.\footnote{In this work we ignore the daily modulation.} $\vmin$ can be found from energy conservation, and is given by \cite{Essig:2015cda}:
\beq\label{vmin}
	\vmin(q,E) = \frac{q}{2 m_\chi} + \frac{E}{q} \ .
\eeq
An useful benchmark for \Eq{diff_rate}, given that the number of semiconductor cells per $\kg$ is $N_{\rm cell} \sim 1 \times 10^{25}$ ($4 \times 10^{24}$) for silicon (germanium), is obtained by taking $20~\MeV$ DM with $\overline{\sigma}_e = 10^{-38}~\cm^2$, which yields a rate of $2 \times 10^4$ ($5 \times 10^3$) events per $\kg\cdot\yr$, with $\kappa$ in \Eq{kappa_correction} giving a $\sim\Ol(1)$ number for typical values of $E$.

\subsection{Astrophysics setup}
\label{subsec:astro_setup}

\Eqs{diff_rate} and (\ref{kappa_correction}) show that the energy and time dependence of the spectrum arises from both the response of the semiconductor to the scattering process and from the DM velocity distribution, encoded in $f_\crs(q, E)$ and $g(\vmin(q, E), t)$ respectively. The latter is given by:
\beqa\label{gvmin}
    g(\vmin, t) & \equiv & \int \!\! \dd v ~ \frac{F(v, t)}{v} \Theta(v-\vmin) \ , \\
    F(v, t) & \equiv & v^2 \!\! \int \!\! \dd \Omega ~ f(\vec{v}; \boldsymbol{\zeta}, \vec{v}_{\rm lab}(t)) \ ,\label{Fv}
\eeqa
with $f(\vec{v}; \boldsymbol{\zeta}, \vec{v}_{\rm lab}(t))$ the normalized distribution of the DM velocity $\vec{v}$ in the lab frame, cut off at the galactic escape velocity which is taken to be $\vesc=528~\km/\seg$ in the galactic rest frame \cite{2019MNRAS.485.3514D}. $\boldsymbol{\zeta}$ are the parameters describing the velocity distribution of the DM components contributing to the local DM. Furtheremore, $\vec{v}_{\rm lab}(t)$ is the velocity of the lab with respect to the galactic rest frame, $v \equiv \vert \vec{v} \vert$ is the DM {\it speed}, $F(v, t)$ the speed distribution, and $\Theta(x)$ the Heaviside step function. The velocity of the lab's frame is given by $\vec{v}_{\rm lab}(t) = \vec{v}_\Sun + \vec{V}_\Earth(t)$, with $\vec{v}_\Sun$ the Sun's velocity in the galactic rest frame, and $\vec{V}_\Earth$ the Earth's velocity in the heliocentric frame. For a detailed list of the numerical values of these astrophysical parameters, see Appendix~\ref{appA}.

As discussed in the Introduction (Sec.~\ref{sec:intro}), stellar substructures due to past mergers have been discovered recently. As a consequence, DM substructures associated with these mergers could contribute to the local DM density, beyond the SHM. These dark substructures will have their own velocity distributions, which will contribute to Eqs.~\eqref{gvmin} and ~\eqref{Fv} as different component terms: $f(\vec{v}) = \sum_i \eta_i f_i(\vec{v})$, $F(v,t) = \sum_i \eta_i F_i(v,t)$, and $g(\vmin, t) = \sum_i \eta_i g_i(\vmin,t)$; where $\eta_i$ is the fraction of the local dark matter that comes from the $i$-th component present. The purpose of the rest of this section is to study the impact of these different velocity distributions on $g(\vmin, t)$, and consequently on the scattering spectrum.

Let us consider a given DM component contributing to the local DM density, with velocity distribution $f(\vec{v}; \boldsymbol{\zeta}, \vec{v}_{\rm lab}(t))$ and corresponding speed distribution $F(v,t)$.\footnote{For brevity we drop the component index $i$. Whether we refer to the total sum of all the components or only to the specific contribution of one of them will be clear from the context.} The effect it will have on $g(\vmin, t)$ can be heuristically understood in terms of three quantities: the value of $v$ at which the component's $F(v,t)$ peaks, which we call the component's {\it most probable speed}; the time of the year at which the DM wind coming from this component is at its largest, called the component's {\it characteristic time} $t_c$; and the {\it coplanarity} $b=\sin \lambda$, where $\lambda$ is the angle between the DM wind and the normal to the Earth's orbital plane: $b=0~(1)$ when the wind is orthogonal (parallel) to the plane \cite{Savage:2006qr, Freese:2012xd}. Note that since $F(v,t)$ is time-dependent, the most probable speed is actually a function of the time of the year as well. Therefore, for convenience we define the most probable speed $\vmp$ as that which maximizes the yearly average $\overline{F}(v)$\footnote{We denote with a bar the yearly average of a time dependent quantity: $\overline{f}(x) = \frac{1}{\rm year} \int\limits_0^{\rm year} \!\! \dd t ~ f(x,t)$.} of the component's speed distribution.

In Table~\ref{table:vmp_t0} we list $\vmp$, $t_c$, and $b$ for possible local DM components we consider throughout this work: the SHM, {\it Gaia}'s halo and Sausage \cite{Necib:2018iwb}, and possible DM streams associated with Nyx \cite{Necib:2019zbk}, S1, S2a, and S2b \cite{OHare:2019qxc} stellar streams. Note that we use the values of the stellar streams as benchmarks for the potential dark matter distributions associated with Nyx, S1, S2a, and S2b. We want to remind the reader again of the caveats which are already mentioned in Sec.~\ref{sec:intro}: we do not claim that these are the true distributions for DM substructures associated with the stellar streams, but only as mere proxies for them. The correlation between stellar streams and their corresponding DM distributions is currently the subject of extensive study. For more details on these DM distributions as well as on how to compute $\vmp$, $t_c$, and $b$, see Appendices~\ref{appA} and~\ref{appB}.

\begin{table}[!ht]
\centering
\def\arraystretch{1.25}
\begin{tabular}{ | c | c | c | c | }
\hline
Components & $\vmp~[\km/\seg]$ & $t_c ~[\text{days}]$ & $b$ \\\hline
SHM & $330$ & $152$ & 0.491 \\[2pt]
{\it Gaia} halo & $304$ & $152$ & 0.491 \\[2pt]
{\it Gaia} Sausage & $259$ & $151$ & 0.477 \\[2pt]
Nyx stream & $192$ & $218$ & 0.860 \\[2pt]
S1 stream & $569$ & $144$ & 0.419 \\[2pt]
S2a stream & $275$ & $358$ & 0.676 \\[2pt]
S2b stream & $227$ & $151$ & 1.00 \\ \hline
\end{tabular}
\caption{$\vmp$, $t_c$, and $b$ for the velocity distributions of the DM components used in this paper. January 1st is day 1. See Appendices~\ref{appA} and~\ref{appB} for details.}
\label{table:vmp_t0}
\end{table}

\subsection{Effects of astrophysics on scattering spectrum}
\label{subsec:astro_spectrum}

Having established the astrophysical setup that we will use throughout this paper, we now devote ourselves to the study of its impact on the DM-$e$ scattering spectrum. We will divide our study into two parts: one dealing with the effects of the most probable speed $\vmp$ on the spectrum's yearly average, and another with the effects of the characteristic time $t_c$ and the coplanarity $b$ on the annual modulation.

\subsubsection{Average Spectrum: the impact of $\vmp$}
\label{subsubsec:average}

Let us begin by considering the impact different DM components have, through their most probable speeds $\vmp$'s, on the yearly average $\overline{g}(\vmin)$. Then we will proceed to consider $\vmp$'s effects on the scattering spectrum via $\kappa$ in Eq.~\eqref{kappa_correction}.

\begin{figure*}[tbh]
    \centering
        \includegraphics[width=0.48\textwidth]{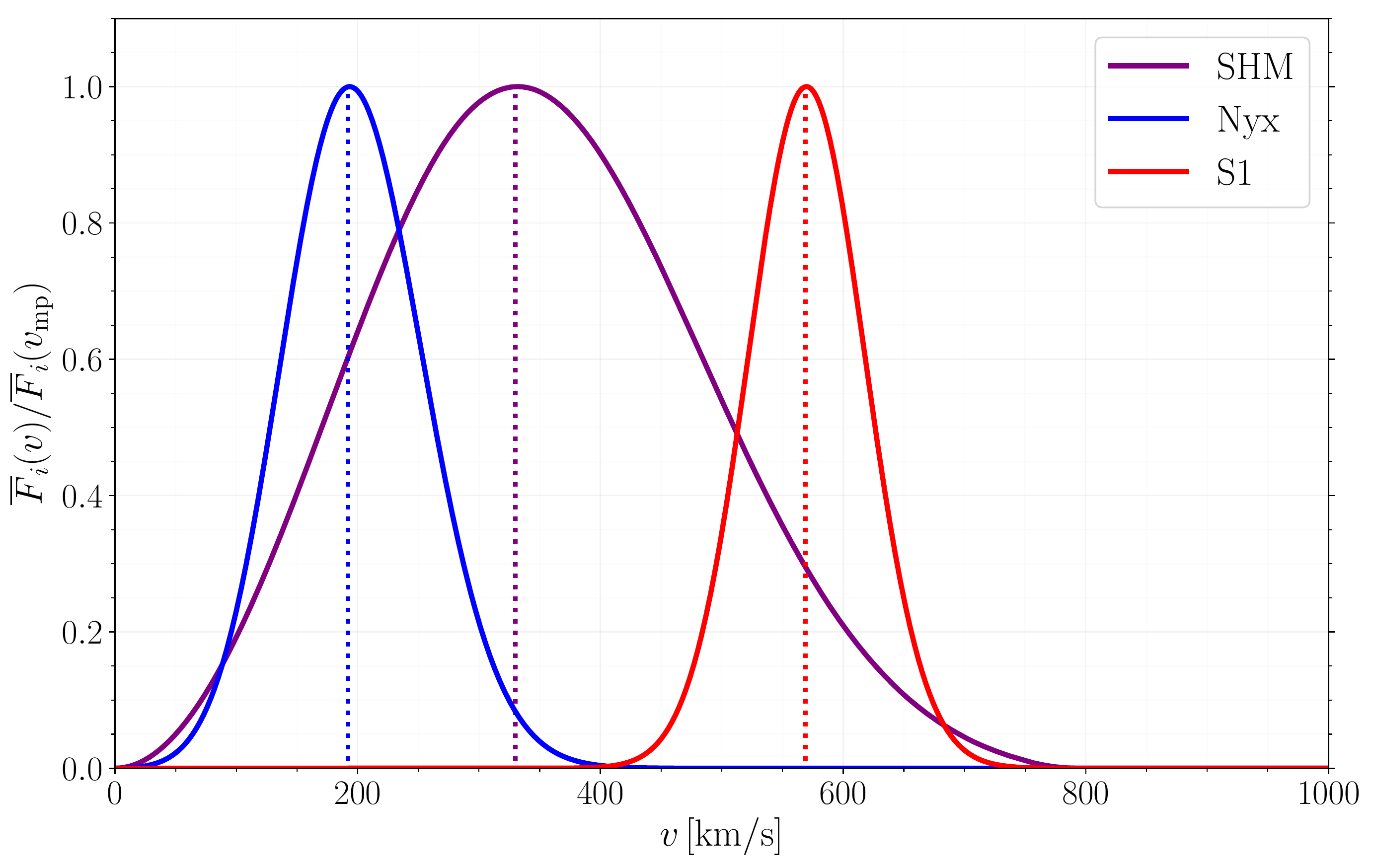}
        \includegraphics[width=0.48\textwidth]{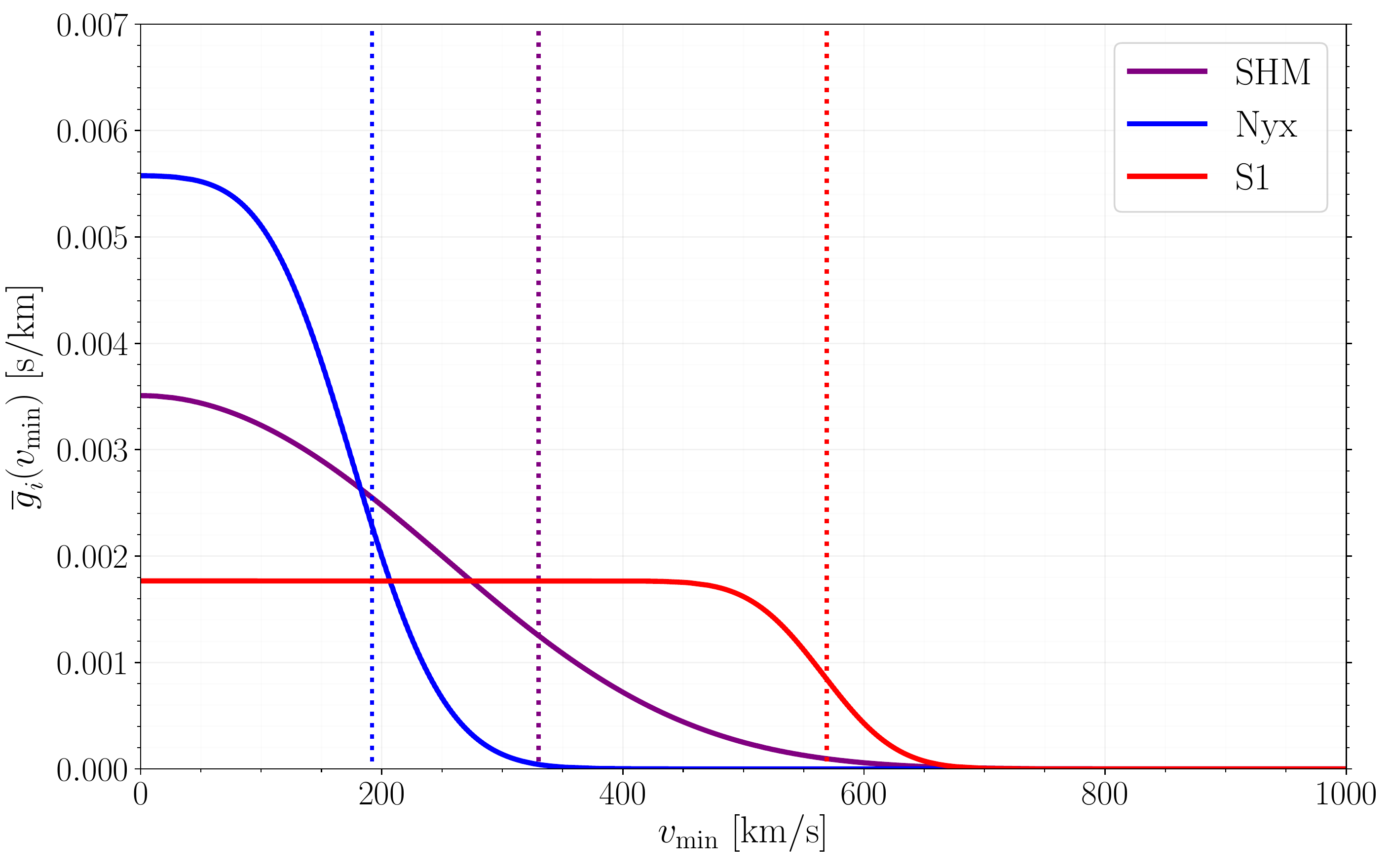}
    \caption{{\it Left:} Yearly average of the DM speed distribution $\overline{F}(v)$ for the SHM (purple) and Nyx (blue) and S1 (red) streams, normalized to their maximums respectively. {\it Right:} Yearly average of the mean inverse speed $\overline{g}(\vmin)$ for the SHM and the Nyx and S1 streams. The vertical dotted lines represent $\vmp$ of each distribution.}
    \label{fig:Fv_gvmin}
\end{figure*}

The left plot in Fig.~\ref{fig:Fv_gvmin} shows $\overline{F}(v)$, normalized to its peak $\overline{F}(\vmp)$, for three example distributions: the SHM in purple, the Nyx stream in blue, and the S1 stream in red. The vertical dotted lines signal the most probable speeds $\vmp$ for each distribution. It can be seen that Nyx, a prograde stream, has a low $\vmp$; whereas S1, a retrograde stream, has a larger one. This is due to the Sun's relative motion with respect to the galactic rest frame.

It is evident from Eq.~\eqref{gvmin} that $g(\vmin, t)$ (as well as its yearly average, $\overline{g}(\vmin)$) is a monotonically decreasing function of $\vmin$. The right panel in Fig.~\ref{fig:Fv_gvmin} plots $\overline{g}(\vmin)$ for the same distributions as before, as well as dotted vertical lines for $\vmin = \vmp$. From the left panel in Fig.~\ref{fig:Fv_gvmin}, we can see that the integrand in Eq.~\ref{gvmin} has most of its support for values of $\vmin < \vmp$, which results in an approximate {\it plateau} at smaller $\vmin$'s for $\overline{g}(\vmin)$. However, for $\vmin > \vmp$, Eq.~\eqref{gvmin} integrates over a diminishing portion of $\overline{F}(v)$, which results in the {\it tail} of $\overline{g}(\vmin)$. We can then speak of a ``width'' for $\overline{g}(\vmin)$, given by $\vmp$. Indeed, comparing S1 with Nyx, we see how $\overline{g}_{\rm S1}(\vmin)$ has support over a wider range of $\vmin$'s than $\overline{g}_{\rm Nyx}(\vmin)$, since $v_{\rm mp, S1} > v_{\rm mp, Nyx}$.

Notice that the maximum height of $\overline{g}(\vmin)$, given by $\overline{g}(0)$ at $\vmin=0$, is inversely correlated with its width. The reason is that $\overline{g}(0)$ is the mean inverse speed of the distribution, which can be related to $\vmp$ as follows:
\beq\label{g0}
    \overline{g}(0) = \left\langle \frac{1}{v} \right\rangle \sim \frac{1}{\langle v \rangle} \sim \frac{1}{\vmp} \ ,
\eeq
where $\langle\cdot\cdot\cdot\rangle$ represents the integral over the speed distribution. Thus, since $v_{\rm mp, S1} > v_{\rm mp, Nyx}$, we have $\overline{g}_{\rm Nyx}(0) > \overline{g}_{\rm S1}(0)$.

Having described the particularities shown by $\overline{g}(\vmin)$ for components of different $\vmp$'s, we now focus on their consequences for $\kappa$, through the $(q, E)$ dependence of $\vmin$ in Eq.~\eqref{vmin}. We also need to inspect the interplay of different factors making up the integrand in Eq.~\eqref{kappa_correction}, which involve not only $\overline{g}(\vmin)$ but also the crystal form factor.

\begin{figure*}[tbh]
    \centering
        \includegraphics[width=0.42\textwidth]{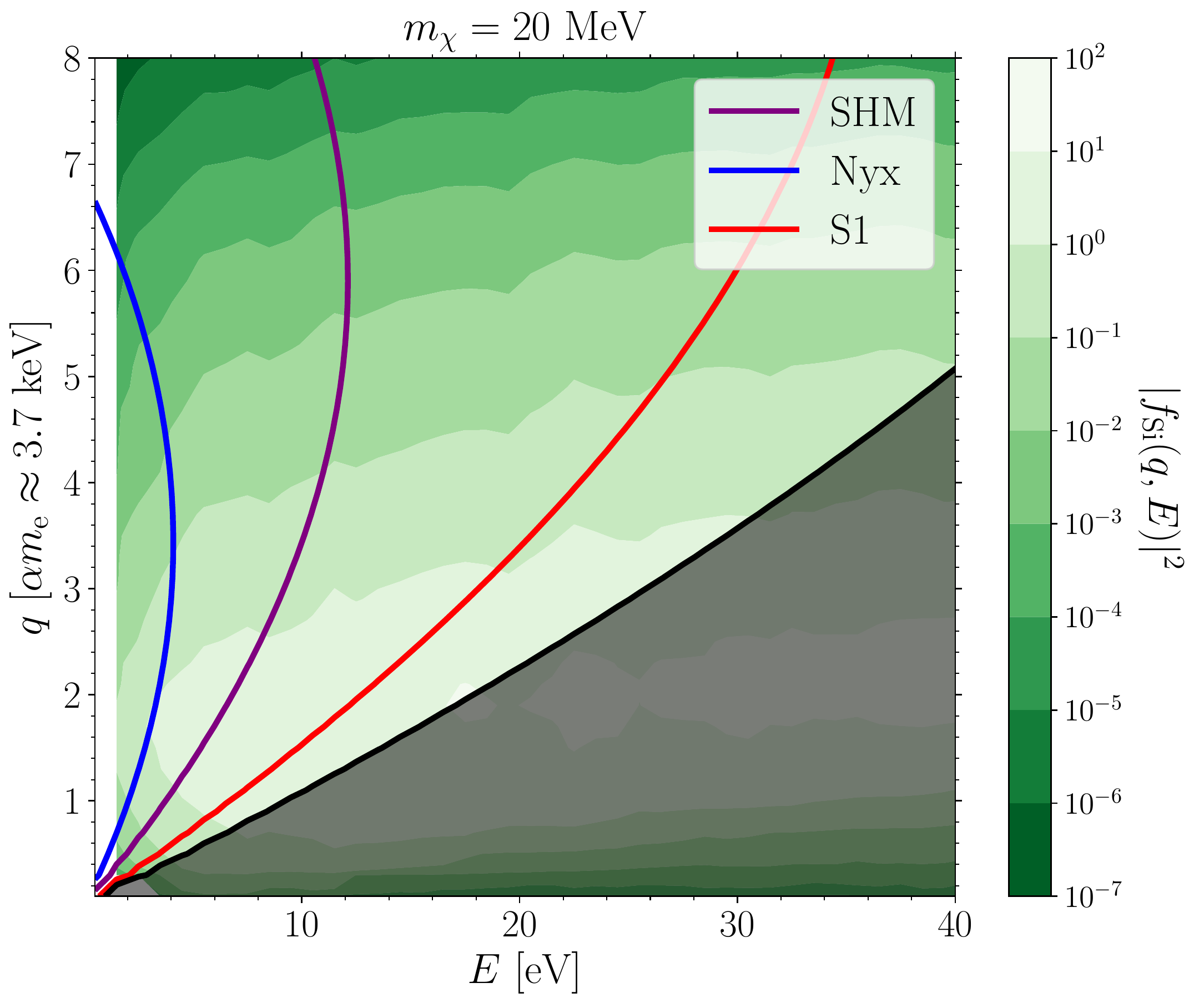}
        \includegraphics[width=0.57\textwidth]{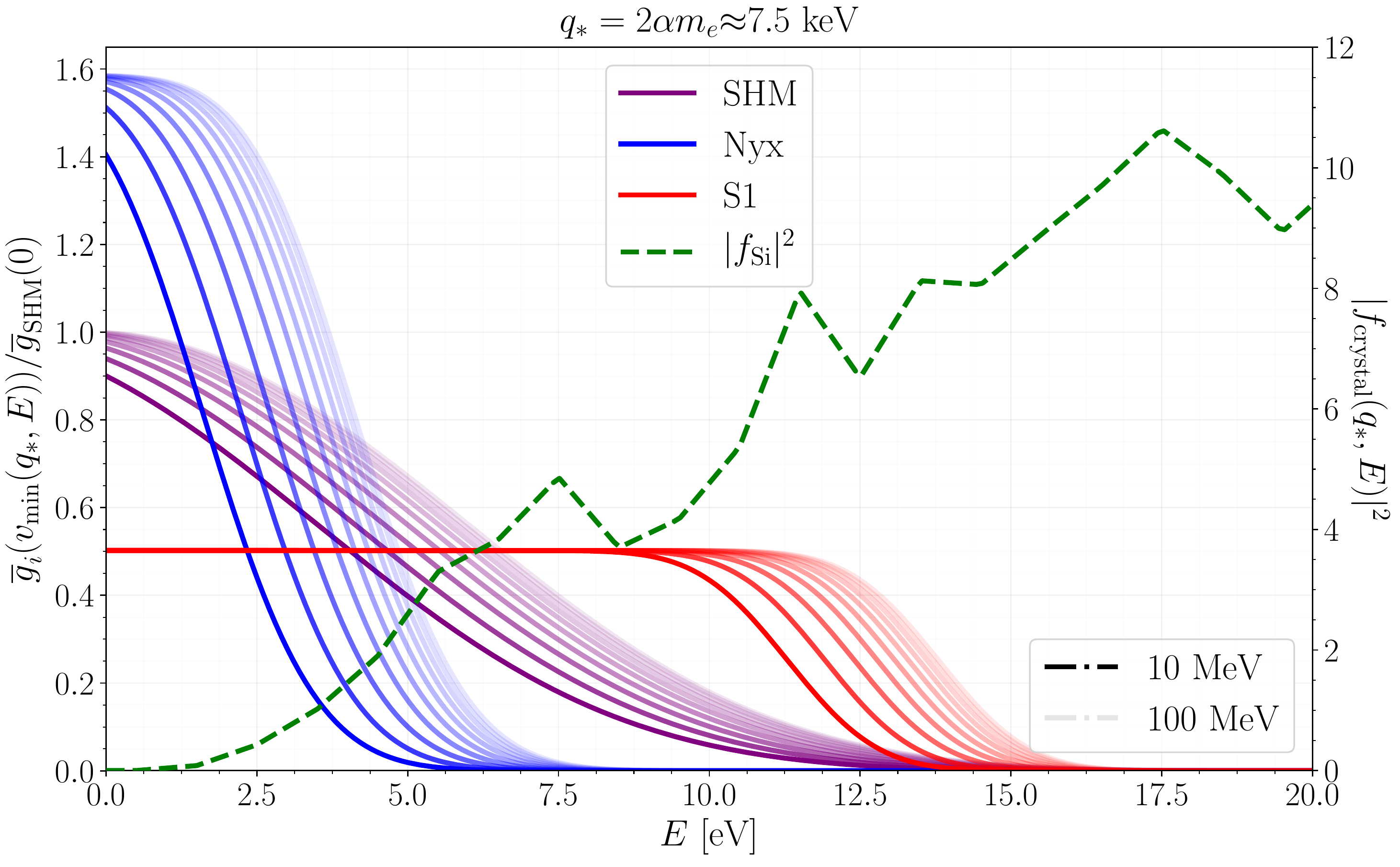}
    \caption{{\it Left:} Contours of the silicon form factor $\vert f_{\rm Si}(q,E) \vert^2$ (green) as a function of $(q, E)$. Also included are the curves in $(q, E)$ space for which $\vmin(q, E) = \vmp$ for SHM (purple), Nyx (blue), and S1 (red), according to Eq.~\eqref{vmin} with $m_\chi = 20~\MeV$. The shaded region corresponds to $\vmin(q,E)$ above the galactic escape velocity (taken to be $\vesc=528~\km/\seg$ in the galactic rest frame), for which there cannot be any scattering events. {\it Right:} $\overline{g}(\vmin(q_*, E))$ at a constant $q_* = 2\alpha m_e$, plotted as a function of $E$ for SHM, Nyx, and S1; and normalized to $\overline{g}_{\rm SHM}(0)$. Decreasing opacity corresponds to increasing DM mass, between $10~\MeV$ and $100~\MeV$. Also plotted in dashed green is the crystal form factor $\vert f_{\rm Si}(q,E) \vert^2$ for silicon.}
    \label{fig:f2_gvmin}
\end{figure*}

The left panel of Fig.~\ref{fig:f2_gvmin} shows the contours of the silicon form factor $\vert f_{\rm Si}(q,E) \vert^2$ as a function of the transferred momentum $q$ and deposited energy $E$.\footnote{The crystal form factors for silicon and germanium, are taken from the publically available tables in \href{http://ddldm.physics.sunysb.edu/ddlDM/}{ddldm.physics.sunysb.edu/ddlDM/}, which were computed with the \href{https://github.com/adrian-soto/QEdark_repo}{{\tt QEdark}} module \cite{Essig:2015cda} of \href{http://www.quantum-espresso.org/}{{\tt Quantum Espresso}} \cite{QE-2009, QE-2017}.} Note that the form factor is at its largest around the typical values of momentum transfer in scattering off electrons: $q\sim {\rm few}\times\alpha m_e$, as well as for energies of order $E\sim{\rm few} \times 10~\eV$. We also present the curves for which $\vmin(q, E) = \vmp$ for SHM, Nyx, and S1, for DM mass $m_\chi=20~\MeV$. To the left of these curves (low $E$) lies the plateau of their corresponding $\overline{g}(\vmin)$; whereas to their right (high $E$) lies its tail. Therefore $\kappa$, and thus the spectrum, decays at large $E$. We could also see from the plot that S1 stream could probe region with larger $E$ with sizable $\vert f_{\rm Si} \vert^2$ while Nyx stream could only probe region with smaller $E$ where $\vert f_{\rm Si} \vert^2$ is suppressed.

From Eq.~\eqref{vmin}, we observe that increasing $m_\chi$ allows for a larger region of $(q, E)$ space to yield sizable values of $\overline{g}(\vmin)$. We show the effects of varying DM mass in the right panel of Fig.~\ref{fig:f2_gvmin}. It shows $\overline{g}(\vmin(q_*, E))$ at constant $q_*=2\alpha m_e$, for SHM, Nyx, and S1, and normalized to its largest SHM value: $\overline{g}_{\rm SHM}(0)$. Also plotted in this panel is the crystal form factors for silicon scaled by 1/10, at the fixed momentum transferred $q_*$. We consider a family of curves with different DM masses between $10$ and $100~\MeV$, with decreasing opacity for larger masses. From the plot, one could see that as expected, when $m_\chi$ increases, $\overline{g}(\vmin(q_*, E))$ has support over larger energies, where the crystal form factors increase. Thus, either larger DM masses or components with larger $\vmp$'s allow for scattering events to occur at larger energies.

In this paper, we focus on silicon target and similar results could be obtained for germanium target as well.

\subsubsection{Annual Modulation: the impact of $t_c$ and $b$}
\label{subsubsec:modulation}

We now consider the time-dependence of the spectrum. As mentioned before, the combined velocities of the Sun around the Milky Way and of the DM particles in a given component result in a {\it ``DM wind''} in the Sun's frame of reference. Since the Earth performs one revolution around the Sun in a year, in the Earth's frame this DM wind displays an annual modulation, which will yield an increase or decrease in the expected number of DM-$e$ events, depending on whether the Earth moves against or with the DM wind, respectively. Throughout the rest of this section, we define modulation as $\delta f(x,t) \equiv f(x,t) - \overline{f}(x)$.

As mentioned in Sec.~\ref{subsec:astro_setup}, a given DM component will have a characteristic time $t_c$. At $t_c$, the most probable speed of the DM particles is at its highest it will be all year; six months later it will be at its lowest. In addition, the component's DM wind will have a coplanarity $b$ with the Earth's orbital plane: maximal coplanarity ($b=1$) will result in the annual modulation having its maximum amplitude, whereas no coplanarity ($b=0$) will result in no modulation at all. Thus, the quantities $t_c$ and $b$ of a given component determine the phase and amplitude of the modulation of the associated DM wind.

\begin{figure*}[tbh]
    \centering
        \includegraphics[width=0.555\textwidth]{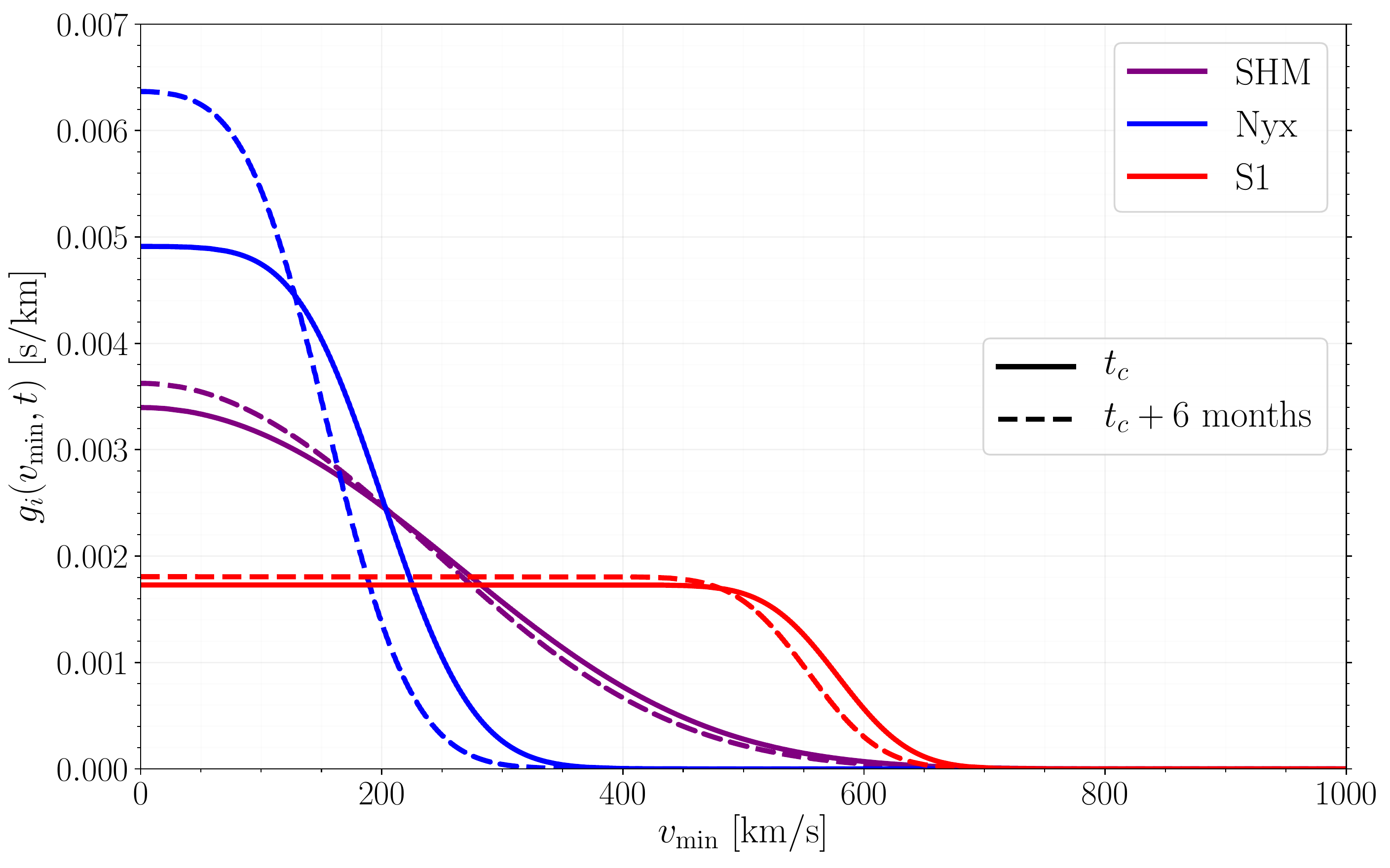}
        \includegraphics[width=0.435\textwidth]{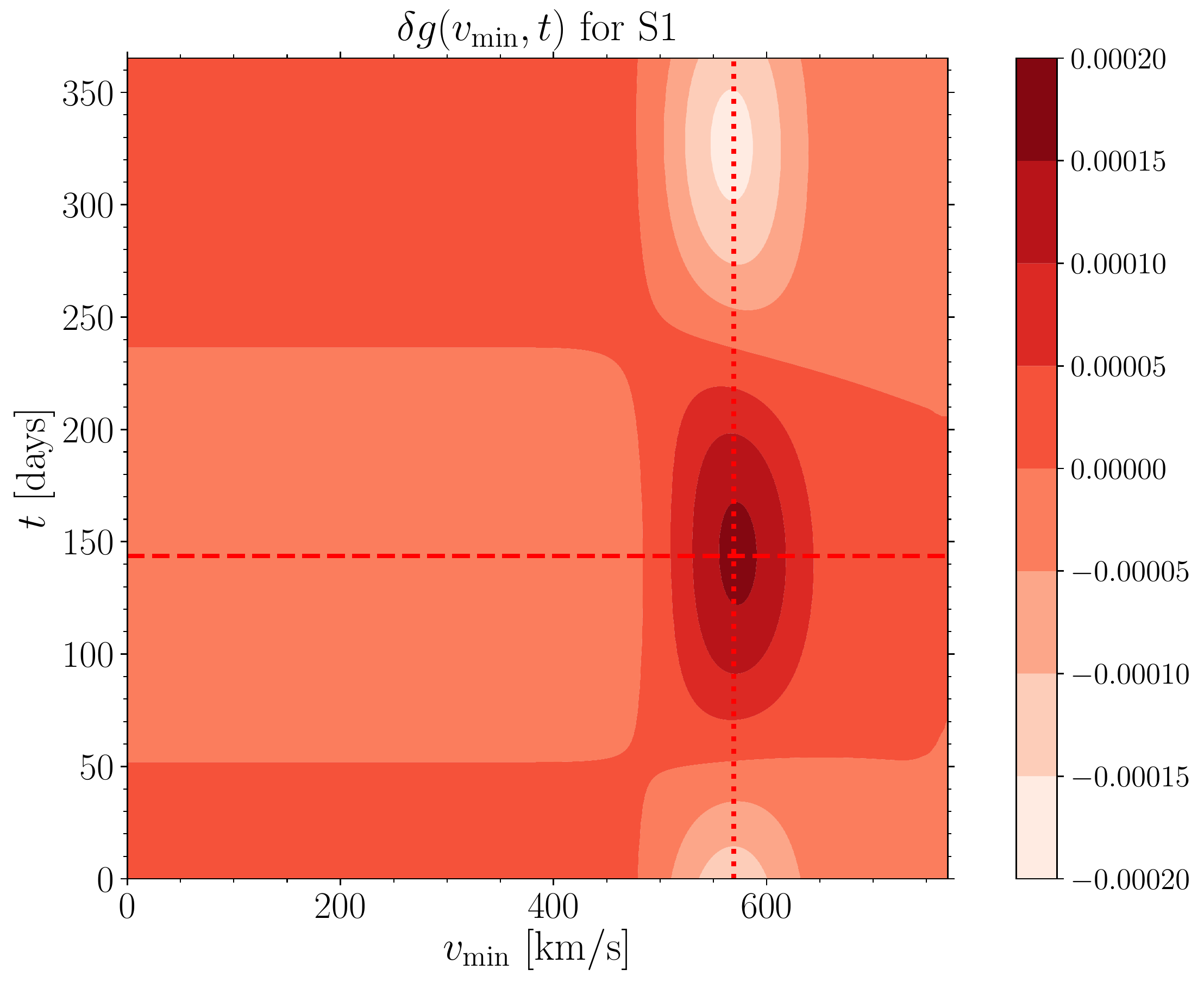}
    \caption{{\it Left:} $g(\vmin, t)$ at times $t_c$ and $t_c+6~{\rm months}$, for SHM (purple), Nyx (blue), and S1 (red). {\it Right:} Example of annual modulation $\delta g(\vmin, t)$ for S1. The dotted line marks $\vmin = v_{\rm mp, S1}$, while the dashed line marks $t=t_{c, {\rm S1}}$.}
    \label{fig:gvmin_tdep}
\end{figure*}

We can use Eq.~\eqref{g0} to understand the effects $t_c$ and $b$ have on $g(\vmin, t)$. As is explained in more detail in Appendix~\ref{appB}, at $t_c$, the relative velocity of the DM wind in the Earth's frame will be at its largest, resulting in a most probable speed given by $\vmp + b V_\Earth$, with $V_\Earth$ the Earth's orbital speed. Six months later the velocity will be at its minimum, given by $\vmp - b V_\Earth$. From Eq.~\eqref{g0} we then arrive at the following expression for the fractional amplitude $A$ of the modulation $\delta g(0, t)$:
\beqa\label{ampl}
	A & \equiv & \frac{\delta g(0, t_c) - \delta g(0, t_c + 6~{\rm months})}{2\overline{g}(0)} \nn \\
	& \approx & -\frac{b V_\Earth}{\vmp} \quad \text{for } V_\Earth \ll \vmp \ .
\eeqa
The reader should keep in mind that Eq.~\eqref{ampl} is only an useful approximation for the magnitude of the modulation amplitude, and strictly speaking, it is valid only for $\vmin = 0$. The negative sign arises from the fact that the height of the plateau for $g(\vmin, t)$, $g(0, t_c)$, is inversely correlated with the most probable speed, so when the latter is at its maximum at $t_c$, $g(0, t_c)$ is at its minimum. The left panel of Fig.~\ref{fig:gvmin_tdep} illustrates this behavior for SHM, Nyx, and S1. Also plotted is $g(\vmin, t_c+6~{\rm months})$, which displays the opposite behavior: the plateau is at its maximum. In addition, the tail of $g(\vmin, t)$ also modulates, but its modulation is out of phase with the plateau's. The reason is that, as discussed in the previous subsection~\ref{subsubsec:average}, the most probable speed is correlated with the width of $g(\vmin, t)$. Thus, when this speed is at its largest at $t_c$, $g(\vmin, t_c)$ is at its widest and it has support for more values of $\vmin$. In summary, plateau and tail of $g(\vmin, t)$ present an annual modulation in opposite ways. The right panel of Fig.~\ref{fig:gvmin_tdep} further illustrates this by showing the contours of the modulation $\delta g(\vmin, t)$ for S1, and marking the time $t=t_c$ and $\vmin=\vmp$. Indeed one could see two opposite modulation phases for $\vmin \lesssim 500$ km/s and $\vmin \gtrsim 500$ km/s.

Finally, since $\vmin$ depends on $E$ (Eq.~\eqref{vmin}), we expect this same behavior to be displayed by the scattering spectrum itself, for low and high values of the energy respectively, in accordance with the right panel of Fig.~\ref{fig:f2_gvmin}. However, there is a subtlety here. For a light DM and a component with a small $\vmp$ such as the Nyx stream, $\vmin$ needed for the scattering above the experimental threshold could always lie in the tail of $g(\vmin, t)$. In this case, we won't observe phase flipping at low and high energies and there will only be one phase observed. For more details, see Appendix~\ref{appC}.

\subsection{Summary}
\label{subsec:summary}

So far we have described how different DM components, characterized by most probable speed $\vmp$, characteristic time $t_c$, and coplanarity $b$, affect the mean inverse speed $g(\vmin, t)$ and consequently the scattering spectrum $\dd R/\dd E$. We have used SHM, Nyx stream, and S1 stream as examples; yet our findings apply to other DM components, and can be summarized as follows:
\begin{itemize}
	\item	$\vmp$ determines the width of $g(\vmin, t)$, as well as the energies $E$ at which the we expect most scattering events.
	\item $\vmp$ is inversely correlated with the height of the plateau of $g(\vmin, t)$, and consequently the number of events at the lowest energies.\footnote{If DM mass is very small, the energy region associated with plateau of $g(\vmin, t)$ may not be kinematically available to the scattering process and only the tail is observed.} The modulation amplitude is also inversely proportional to $\vmp$.
	\item $t_c$ determines the phase of the annual modulation of both $g(\vmin, t)$ and the scattering spectrum: the time of the year at which the plateau (tail) is minimized (maximized).
	\item $b$ determines the amplitude of the annual modulation.
\end{itemize}

In an experiment such as SENSEI \cite{Abramoff:2019dfb, Barak:2020fql} or EDELWEISS \cite{Arnaud:2020svb} the scattering spectrum is observed not as a continuum in $E$, but as a function of the number $Q$ of electron-hole pairs detected in the semiconductor, also called the {\it ionization} level of the semiconductor. A very simple map between $E$ and $Q$ is given by \cite{Essig:2015cda}:
\beq\label{Qdef}
	Q = \left( 1 + \left\lfloor \frac{E - E_{\rm gap}}{\varepsilon} \right\rfloor \right) \Theta(E - E_{\rm gap}) \ ,
\eeq
where $E_{\rm gap}$ is the band-gap energy of the semiconductor and $\varepsilon$ the mean energy per electron-hole pair. $E_{\rm gap} = 1.2~\eV$ and $\varepsilon = 3.8~\eV$ for silicon, while $E_{\rm gap} = 0.67~\eV$ and $\varepsilon = 2.9~\eV$ for germanium.

\begin{figure*}[tbh]
    \centering
        \includegraphics[width=0.48\textwidth]{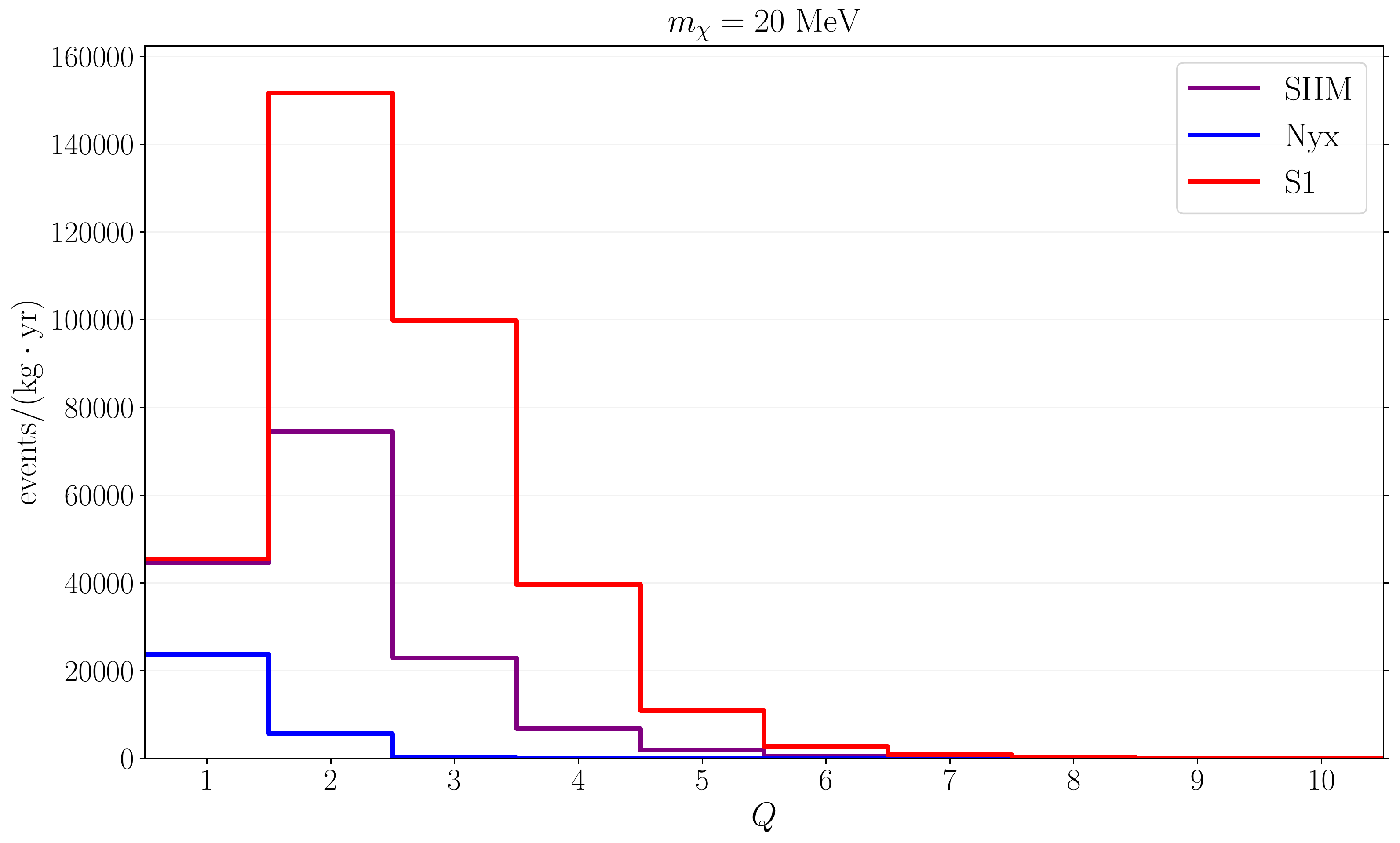}
        \includegraphics[width=0.48\textwidth]{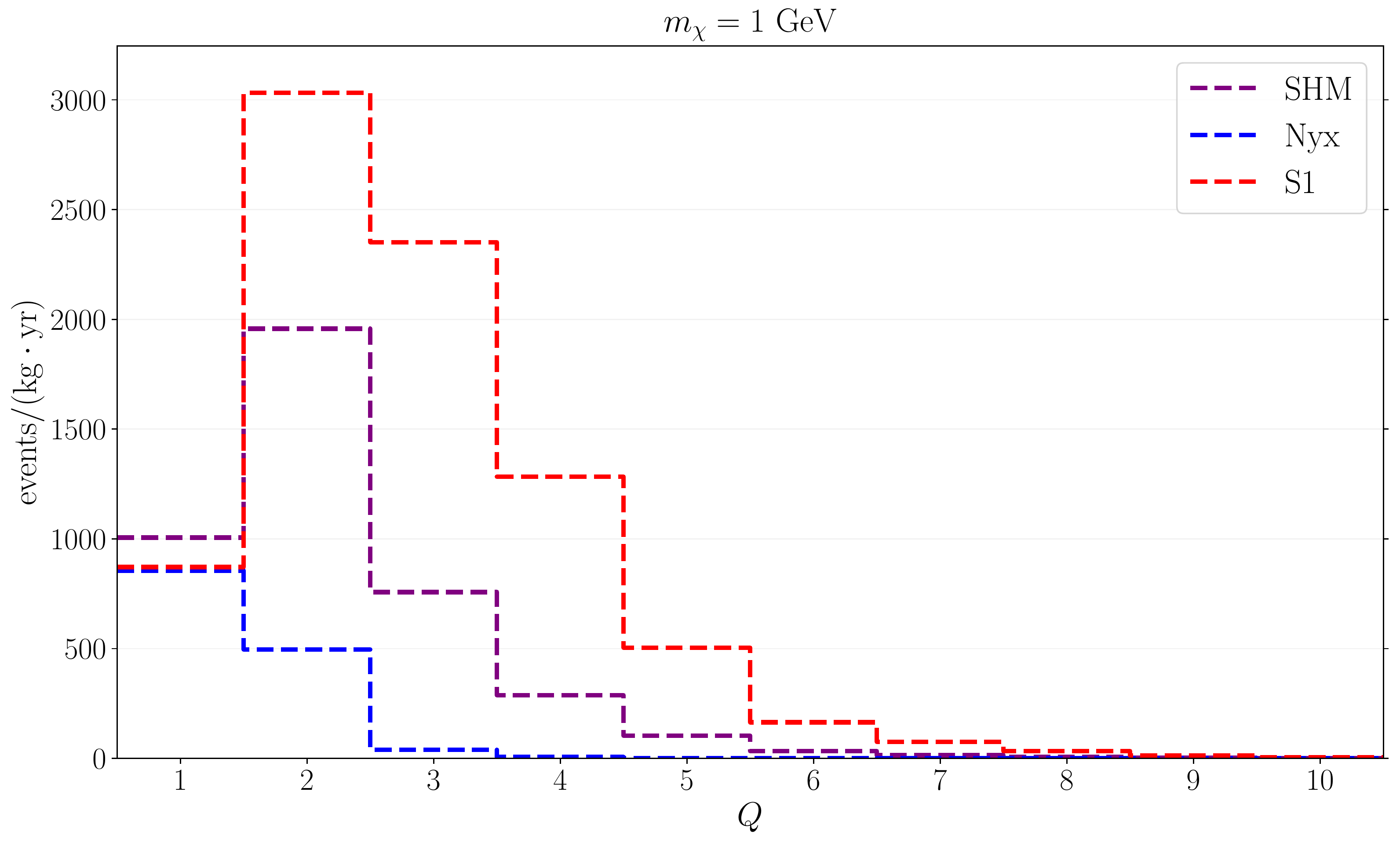}
    \caption{$Q$-binned scattering spectrum off silicon, per $\kg\cdot\yr$ of exposure, for DM mass $m_\chi=20~\MeV$ ({\it left}) and $m_\chi=1~\GeV$ ({\it right}); fixing $F_\DM = 1$, $\overline{\sigma}_e = 10^{-37}~\cm^2$, and assuming all DM particles coming from SHM (purple), Nyx (blue), or S1 (red).}
    \label{fig:1d_rates}
\end{figure*}

\begin{figure*}[tbh]
    \centering
        \includegraphics[width=0.48\textwidth]{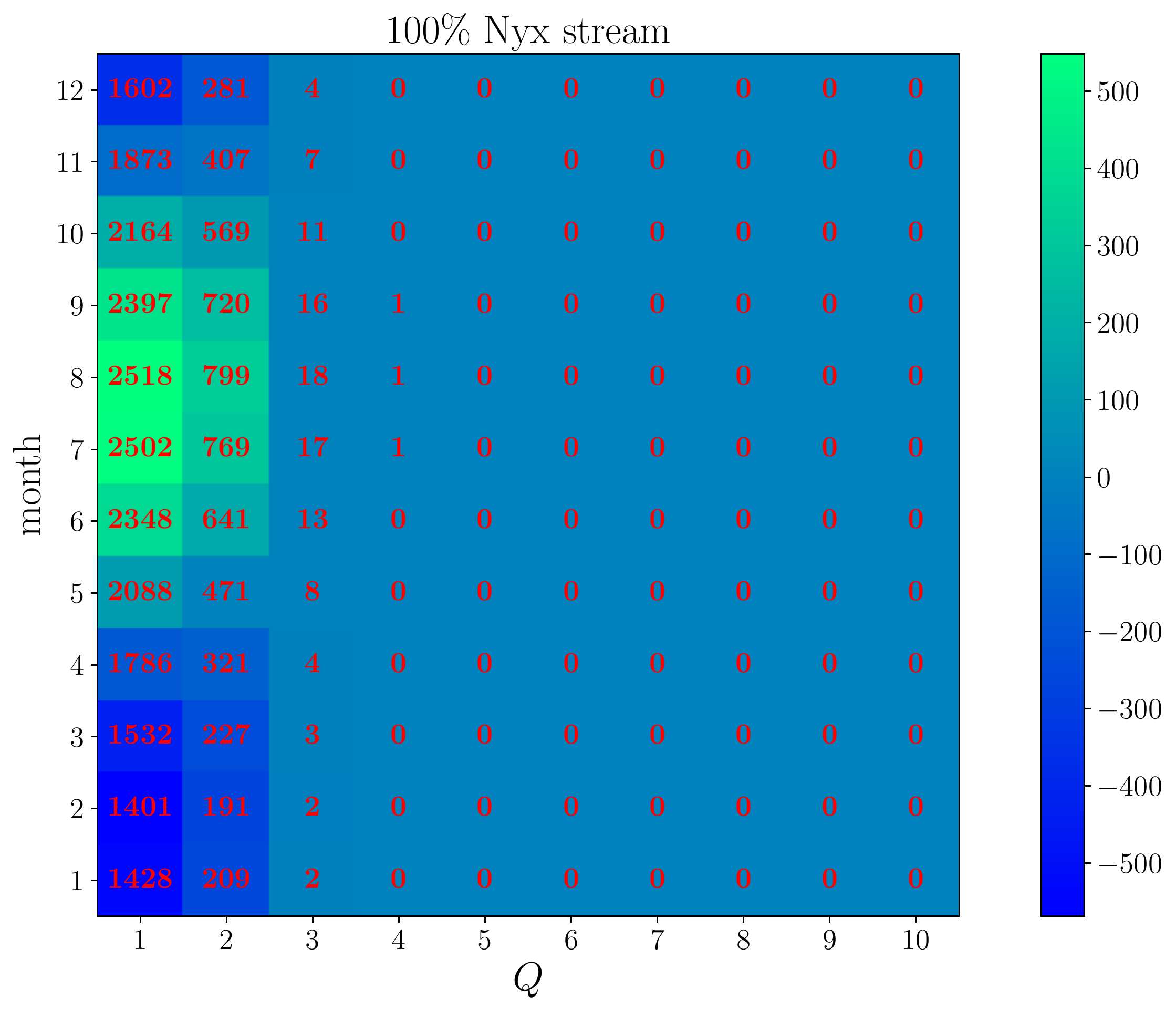}
        \includegraphics[width=0.48\textwidth]{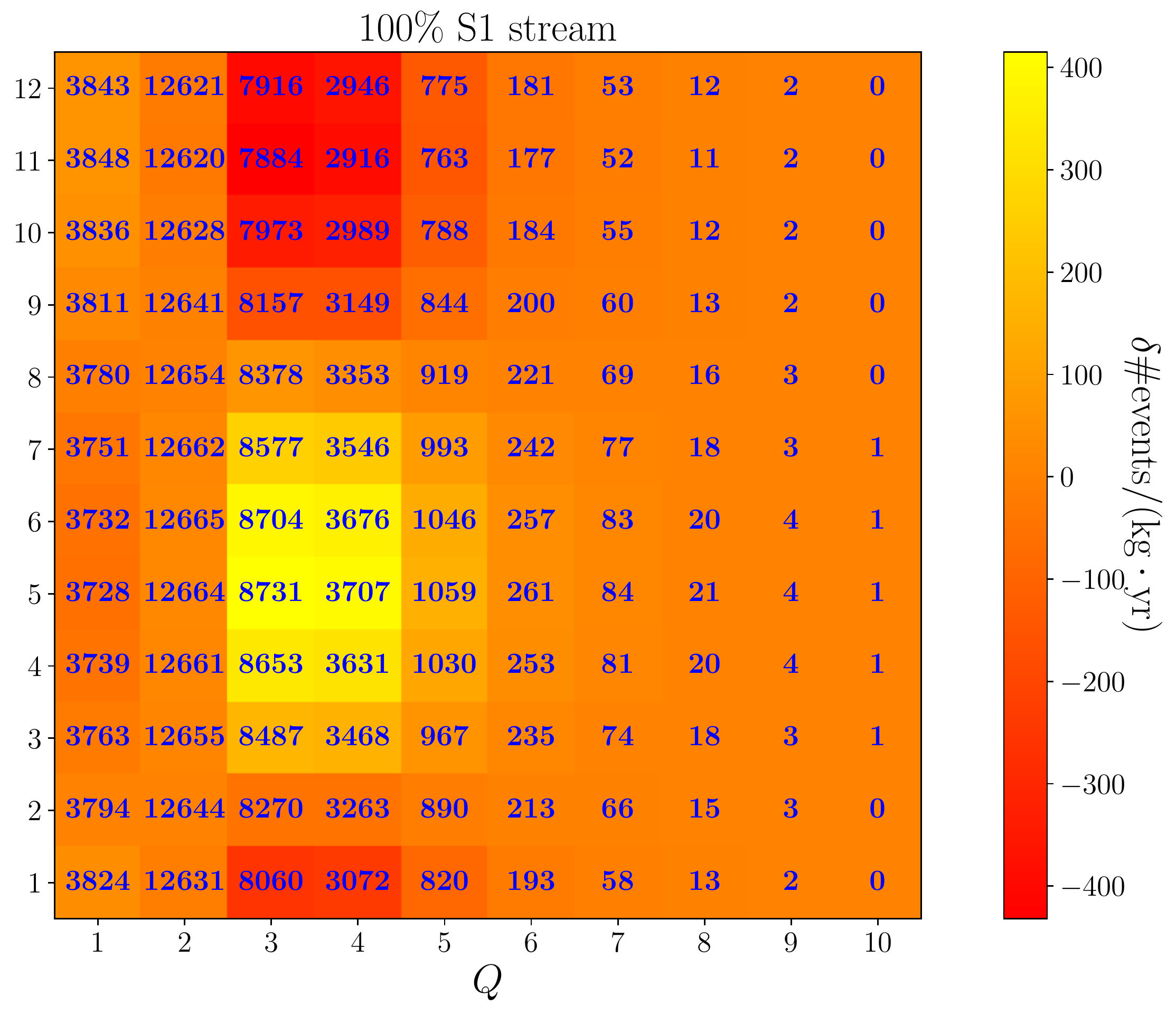}
    \caption{$Q$-month binned scattering spectra off silicon for DM mass $m_\chi=20~\MeV$, $F_\DM = 1$, $\overline{\sigma}_e = 10^{-37}~\cm^2$, assuming that all DM particles coming from the Nyx stream ({\it left}), or the S1 stream ({\it right}). The numbers indicate the expected number of events in that bin, while the colors correspond to the annual modulation and indicate whether the numbers are above or below the yearly average.}
    \label{fig:heatmaps}
\end{figure*}

Experimentally there is then a natural binning of the $E$ axis in terms of $Q$. In Fig.~\ref{fig:1d_rates} we show the yearly average of this binned energy spectra for $m_\chi=20~\MeV$ and $1~\GeV$, for the extreme cases where 100\% of the local DM comes from SHM, Nyx, or S1 components.
Binning the $t$ axis as well, for example in months, we can describe the scattering spectrum, for different combinations of DM and astrophysical parameters, as the expected number of events in a time-energy bin $(t_i, Q_j)$. Fig.~\ref{fig:heatmaps} shows the $Q$-month binned scattering spectrum off silicon for DM of mass $m_\chi=20~\MeV$, $n=0$ ($F_\DM = 1$), and $\overline{\sigma}_e = 10^{-37}~\cm^2$ per $\kg-\yr$ of exposure, for 100\% Nyx or S1 components. Both Fig.~\ref{fig:1d_rates} and Fig.~\ref{fig:heatmaps} confirm the relations we list above.

\section{Analysis setup}
\label{sec:statistics}

So far we have only focused on the phenomenology of DM-$e$ scattering with semiconductor targets for several DM substructures. At this stage, we can pose a couple of legitimate questions: {\it a)} what are the prospects of detecting a DM signal with next-generation electron recoil experiments for a given DM velocity distribution? {\it b)} assuming discovery, can we distinguish the effects of DM substructures such as streams in a statistically significant way? While part {\it a)} has received attention in the literature, most notably in the pioneering work by Refs.~\cite{Essig:2015cda, Lee:2015qva} (see also ref.~\cite{Andersson:2020uwc} for a recent exploration), a detailed examination of part {\it b)} in context of DM-$e$ scattering experiments, to the best of our knowledge, has not been performed yet. With that in mind, we outline a statistical analysis in this section that uses forecasting as a complementary probe of DM substructure properties alongside the usual discovery reach contours.

\renewcommand{\tabcolsep}{1.5mm}
\renewcommand\arraystretch{1.8}
\begin{table}[ht]
	\centering
	\begin{tabular}{| c | c  | c	 | c  |}
	\hline
	 Background model & $R^{\rm bkg}_{Q =1}$ &  $R^{\rm bkg}_{Q =2}$ & $R^{\rm bkg}_{\rm const}$ \\ \hline
	 Maximum (MAX)~\cite{Barak:2020fql} & $\sim 10^8$ & $\sim 10^6$ & $\sim 10^3$ \\[2pt]
	 Optimistic (OPT)~\cite{Tiffenberg2018} &  -- & -- & $\sim 1$ \\ \hline
     \end{tabular}
\caption{Approximate background rates for two configurations of an idealized SENSEI-like experiment with Si target~\cite{Tiffenberg:2017aac} in units of events/kg/day. The OPT model has an ionization threshold $Q_{\rm th}=3$ (hence no entries for $R^{\rm bkg}_{Q =1, 2}$) to mitigate the background through sideband measurements. Following~\cite{Tiffenberg2018}, the constant background rate is taken to be uniform across all bins.} \label{tab:dd_summary}
\end{table}

However, an important complication in detecting DM and estimating its properties with electron recoil data is the existence of experimental background due to ``dark current" events. Frustratingly, the background, along with the DM signal for a large range of masses and DM form factors, peaks at low ionization $Q$ bins~\cite{Abramoff:2019dfb, Aguilar-Arevalo:2019wdi, Barak:2020fql}. Given that, we also consider a more optimistic possibility of mitigating a large background component through sideband measurements, which can be achieved for recoil data with higher thresholds $Q_{\rm th}$.\footnote{R. Essig, private communication.} We also propose the use of time domain data to probe the characteristic modulating component of a potential DM signal. Since we can reasonably assume the background to be time-independent, the time domain channel can essentially be treated as background-free for downstream analysis after DM discovery. Although we will derive our results in this paper assuming an idealized SENSEI-like~\cite{Tiffenberg:2017aac} experiment outlined in Table~\ref{tab:dd_summary}, the formalism described here may easily be extended to experiments with different targets.

Thus, in the rest of the section, we develop a profile likelihood analysis applicable to next-generation electron recoil experiments with two main goals:
\begin{itemize}
\item estimate the discovery reach for: {\it i)} a DM signal in presence of a realistic experimental background, and {\it ii)} an annually modulating DM signal over the average recoil spectrum by leveraging full time domain data.

\item constrain the DM fraction in substructure by distinguishing signals from various astrophysical configurations.
\end{itemize}
For ease of reference, we summarize the important details of our statistical analysis in Table~\ref{tab:stat_summary}.

\renewcommand{\tabcolsep}{1.5mm}
\renewcommand\arraystretch{1.8}
\begin{table}[ht]
	\centering
	\begin{tabular}{| c | c  | c	 |}
	\hline
	 Statistical test & DM parameters & Relevant eqs. \\ \hline
	 DM signal discovery & $\bar{\sigma}_e$, $m_\chi$ & \eqref{eq:likelihood_1}, \eqref{eq:likelihood_2}, \eqref{eq:ds_reach} \\[2pt]
	 Modulation discovery & $\bar{\sigma}_e$, $m_\chi$ & \eqref{eq:likelihood_1}, \eqref{eq:ds_reach}, \eqref{eq:likelihood_3} \\[2pt]
	 Sensitivity forecasts & $\eta$, $m_\chi$ & \eqref{eq:likelihood_1}, \eqref{eq:likelihood_2}, \eqref{eq:mod_compare} \\ \hline
     \end{tabular}
\caption{Summary of different methods and its ingredients, namely the DM signal parameters that we will vary and relevant formulae, used in our statistical analysis.} \label{tab:stat_summary}
\end{table}

We define the likelihood function (henceforth simply the likelihood) for a hypothetical experiment which detects electron-hole pairs produced from a DM-$e$ recoil event in both ionization and time bins,
\beq\label{eq:likelihood_1}
\mathcal{L}(\mathcal{D}|\bm \psi) \equiv \prod_{i= 1}^{n_t} \prod_{j= 1}^{n_Q} \mathcal{L}_b ( N_{\rm obs} (t_i, Q_j) \, | \, \bm \psi),
\eeq
where the product is over both time ($n_t$) and ionization ($n_Q$) bins respectively. The likelihood $\mathcal{L}_b(N|\boldsymbol{\psi})$ in general (we note an exception later in the section) is given by the Poisson probability distribution,
\beq \label{eq:likelihood_2}
\begin{split}
\loge \, \mathcal{L}_b(N_{\rm obs} \, | \, \bm \psi)  & = N_{\rm obs}  \cdot \loge \, N_{\rm th} ((\sig(t_i, Q_j) \, | \, \boldsymbol{\theta}, \boldsymbol{\lambda} \, ),  \, \bkg) \, \\
& - N_{\rm th}  ((\sig(t_i, Q_j) \, | \, \boldsymbol{\theta}, \boldsymbol{\lambda} \, ), \, \bkg),
\end{split}
\eeq
such that $N_{\rm obs}$ and $N_{\rm th}$ are the number of observed and predicted events in the $i^{\rm th}$ time and $j^{\rm th}$ ionization bin. For brevity, we have dropped all constant terms from the expression above. The predicted events in each bin consist of the signal rate $S_{ij}(\boldsymbol{\theta}, \boldsymbol{\lambda})$, the background rate $B_{i j}$ along with an overall normalization given by the exposure $E$. Assuming a linear model, we have
\beq \label{eq:lin_model}
N_{\rm th} ((\sig (t_i, Q_j) \, | \, \boldsymbol{\theta}, \boldsymbol{\lambda} \,), {\bf B})  = (S_{ij} + B_{ij}) \cdot E.
\eeq
The signal is evaluated for each DM model, ${\bm \psi} = (\boldsymbol{\theta}, \boldsymbol{\lambda} \,)$, where $\boldsymbol{\theta}$ and $\boldsymbol{\lambda}$ are the signal and nuisance parameters respectively. The identification of a model parameter as a signal or nuisance parameter is determined by the nature of the analysis.\footnote{For example, we treat the fraction of DM in substructure as a nuisance parameter while estimating the discovery reach in $\bar{\sigma}_e - m_\chi$ space, but as a signal parameter while constraining DM's astrophysical properties. Furthermore, for the case of an unknown background, ${\bf B}$ should also be treated as a nuisance parameter.} Lastly, we note that the definitions above can be trivially extended to the average spectrum case by dropping all $t$ bins and using only the $Q$ bins with the recoil rate derived for the mean Earth velocity. In experimental terms, this is equivalent to working with all the data collected through the duration of the experiment in each $Q$ bin.

Before formally defining a test statistic (TS) for each of the goals we stated above, we note that in the absence of {\it real} data, we use the Asimov data set~\cite{Cowan:2010js} for estimating the median sensitivity of an idealized next-generation experiment. In practice, the Asimov data set is simply the {\it mock} signal (plus background wherever applicable) corresponding to the parameter values of a chosen benchmark point with no statistical fluctuations.

The discovery reach for an experiment is expressed through the likelihood ratio TS~\cite{1933RSPTA.231..289N},
\beq \label{eq:ds_reach}
q_0 = - 2 \, {\rm ln} \, \left(\frac{\mathcal{L}(\mathcal{D}_{\rm Asm} \, | \, \boldsymbol{\theta} = 0)}{\mathcal{L}(\mathcal{D}_{\rm Asm} | \, \boldsymbol{\theta} \, )}\right) \sim \chi^2_1,
\eeq
where $\mathcal{D}_{\rm Asm}  \equiv N_{\rm Asm} (\boldsymbol{\theta}, \boldsymbol{\lambda} \,) $ such that $(\boldsymbol{\theta}, \boldsymbol{\lambda} \,)$ correspond to the signal and nuisance parameter values respectively for the chosen benchmark point. The ratio can be shown to have a $\chi^2$ distribution with 1 degree of freedom in the asymptotic limit~\cite{Wilks:1938dza, Cowan:2010js}, whereas the significance $Z$ of the detection is then expressed in terms of the inverse CDF of the normal distribution $\Phi(x)$ for a given $p$-value,
\beq \label{eq:significance}
Z = \Phi^{-1}(1 - p).
\eeq

For the case of DM discovery, Eq.~\eqref{eq:ds_reach} can be evaluated in a straightforward fashion by substituting the Poisson likelihood for the average recoil spectrum given by Eq.~\eqref{eq:likelihood_2} into Eq.~\eqref{eq:likelihood_1}. However, for the discovery of an annually modulating signal, the likelihood must take into account that the modulation, defined as the bin-by-bin difference between the time domain and average rates, is not an experimental observable. Instead, we note that since both the rates are Poisson distributed individually, their difference follows the Skellam distribution given by,
\beq\label{eq:likelihood_3}
\begin{split}
\mathcal{L}_b( N_{\rm{mod}} & (t_i, Q_j)  | N_{\rm tim}, N_{\rm avg} ; \bm \psi) \equiv e^{-(N_{\rm tim} + N_{\rm avg})}  \\
& \times \, \left(\frac{N_{\rm tim}}{N_{\rm avg}} \right)^{N_{\rm mod}/2} \, I_{N_{\rm mod}}(2 \sqrt{N_{\rm tim} N_{\rm avg}}),
\end{split}
\eeq
where $N_{\rm{mod}}$ is the number of modulation events in the $i^{\rm th}$ time and $j^{\rm th}$ ionization bin and may be positive or negative, depending on the values of $N_{\rm tim}$ and $N_{\rm avg}$, respectively the mean number of time domain and average events; $I_k(x)$ is the modified Bessel function of the first kind. Thus, for estimating the discovery reach of a modulating signal, we adopt the TS in Eq.~\eqref{eq:ds_reach} with the likelihood function defined in Eq.~\eqref{eq:likelihood_3}.

Meanwhile, we use sensitivity forecasts to illustrate an experiment's capability to distinguish signals from various DM substructure components. The TS, based on the pairwise comparison of neighboring parameter points, is defined as,
\beq \label{eq:mod_compare}
{\rm TS} = - 2 \, {\rm ln} \, \left(\frac{\mathcal{L}(\mathcal{D}_{\rm Asm} (\boldsymbol{\theta}_2) \, | \, \boldsymbol{\theta}_1)}{\mathcal{L}(\mathcal{D}_{\rm Asm} (\boldsymbol{\theta}_2) \, | \, \boldsymbol{\theta}_2)}\right) \sim \chi^2_1,
\eeq
where the Asimov data $\mathcal{D}_{\rm Asm}$ is defined analogous to that in the discovery reach case assuming the Poisson likelihood in Eq.~\eqref{eq:likelihood_2}. In frequentist statistical terms, the TS above is used to reject the null hypothesis that signals corresponding to $\boldsymbol{\theta}_1$ and $\boldsymbol{\theta}_2$ are indistinguishable at the $(1 - \alpha) \, \%$ confidence level (CL). We use the euclideanized signal (ES) method introduced by Refs.~\cite{Edwards:2017mnf, Edwards:2017kqw} for a fast, benchmark-free calculation of Eq.~\eqref{eq:mod_compare}. For more details on how the ES method is implemented in context of direct detection experiments, we refer the reader to Refs.~\cite{Edwards:2018lsl, Buch:2019aiw}.


\section{Results}
\label{sec:results}

In this section, we present the results of our statistical analysis in form of discovery reaches and sensitivity forecasts. We focus on the potential of an idealized SENSEI-like electron recoil experiment for DM discovery, and for probing the fraction of the local DM density, $\eta$, in kinematic substructure such as a stream and a debris flow using the observed spectrum. We use the velocity distributions described in Sec.~\ref{subsec:astro_setup} as benchmarks for a phenomenological study, and treat DM mass, cross section, and DM substructure fraction(s) as free parameters depending on the analysis. All our results have been derived assuming a 1 kg-year exposure and background models summarized in Table~\ref{tab:dd_summary}. In Appendix~\ref{appD}, we include additional supplemental results.

\subsection{Discovery reaches} \label{sec:discreaches}
\begin{figure*}[tbh]
    \centering
        \includegraphics[width=0.48\textwidth]{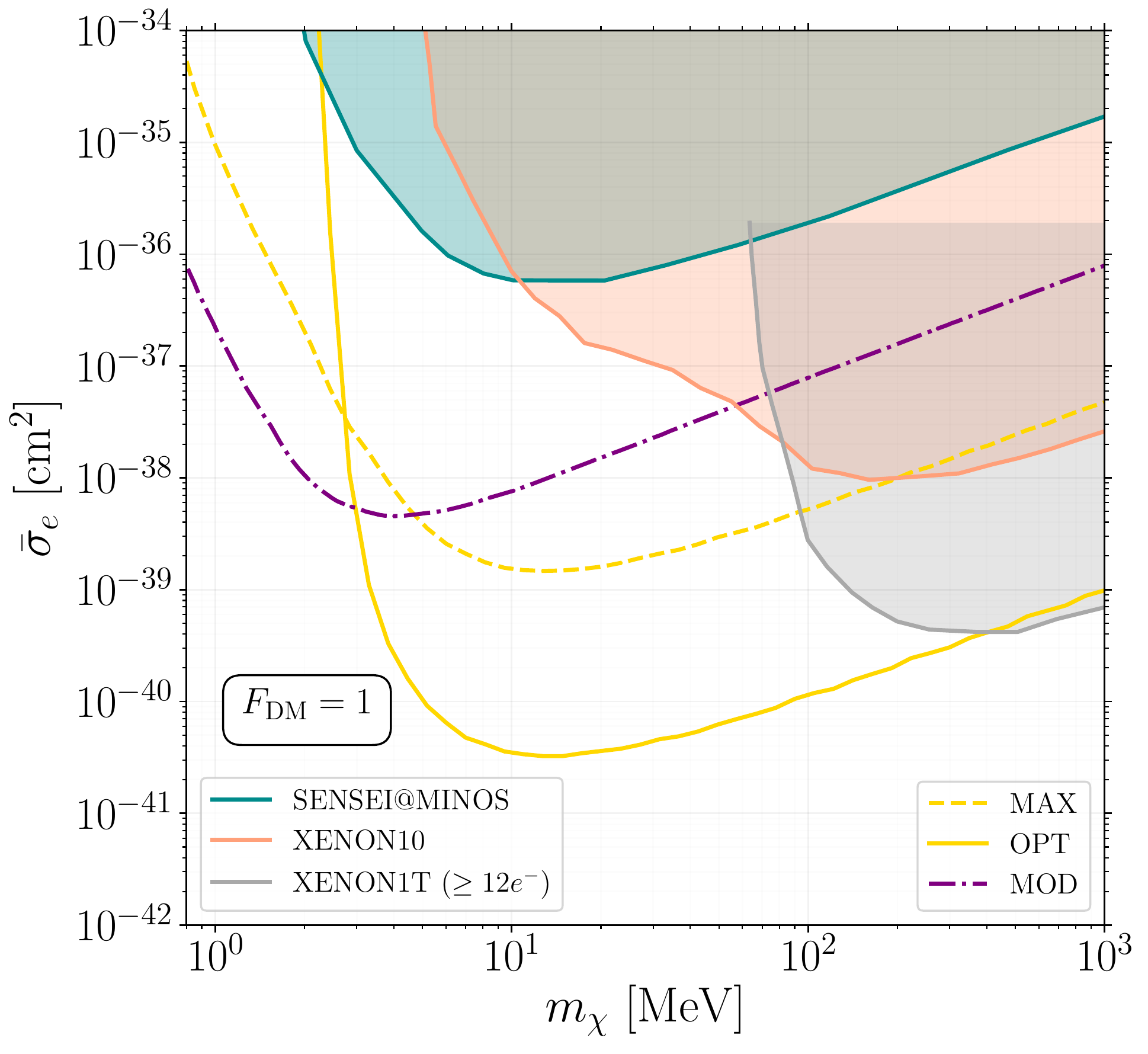}
         \includegraphics[width=0.48\textwidth]{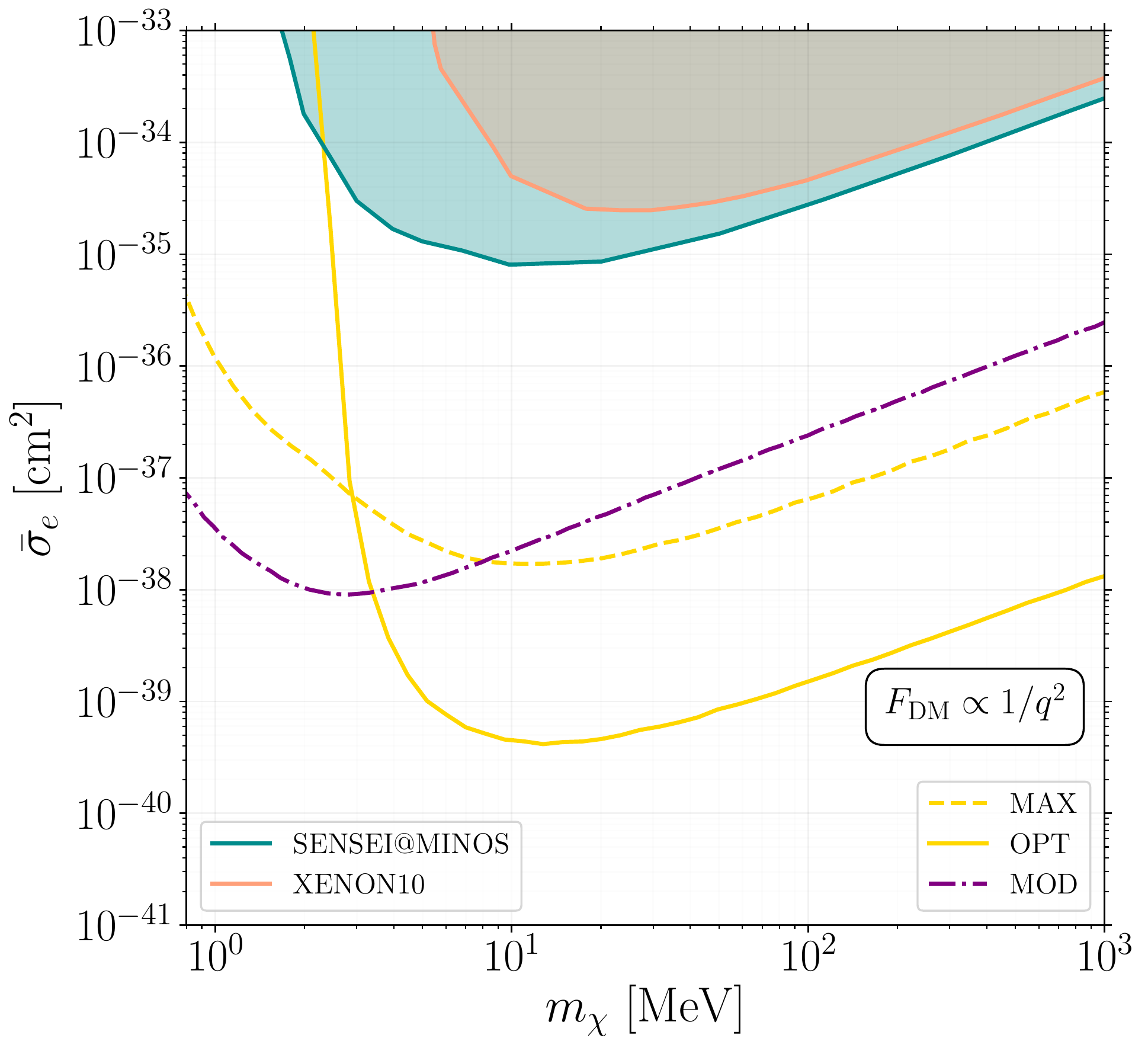}
         \includegraphics[width=0.48\textwidth]{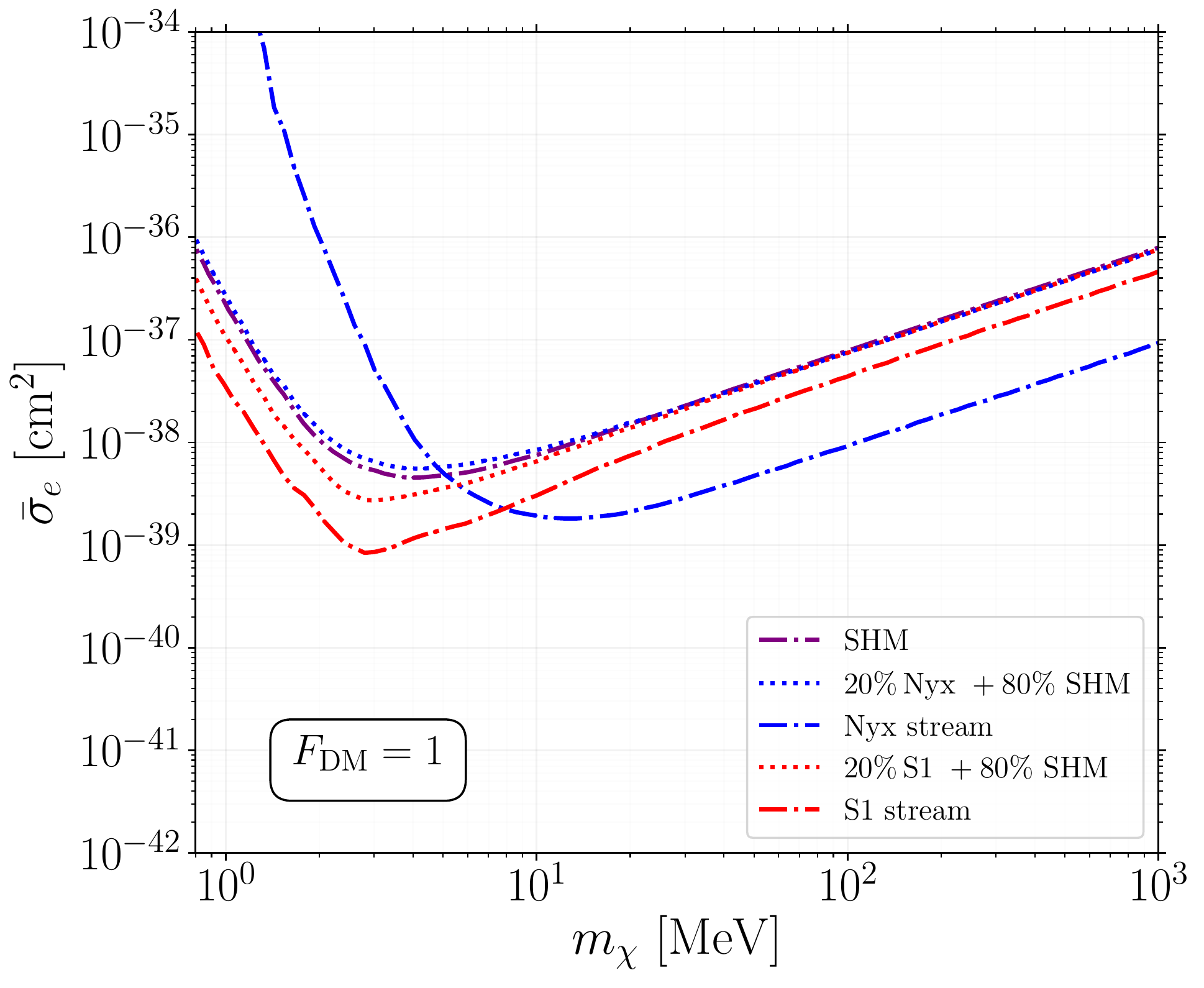}
         \includegraphics[width=0.48\textwidth]{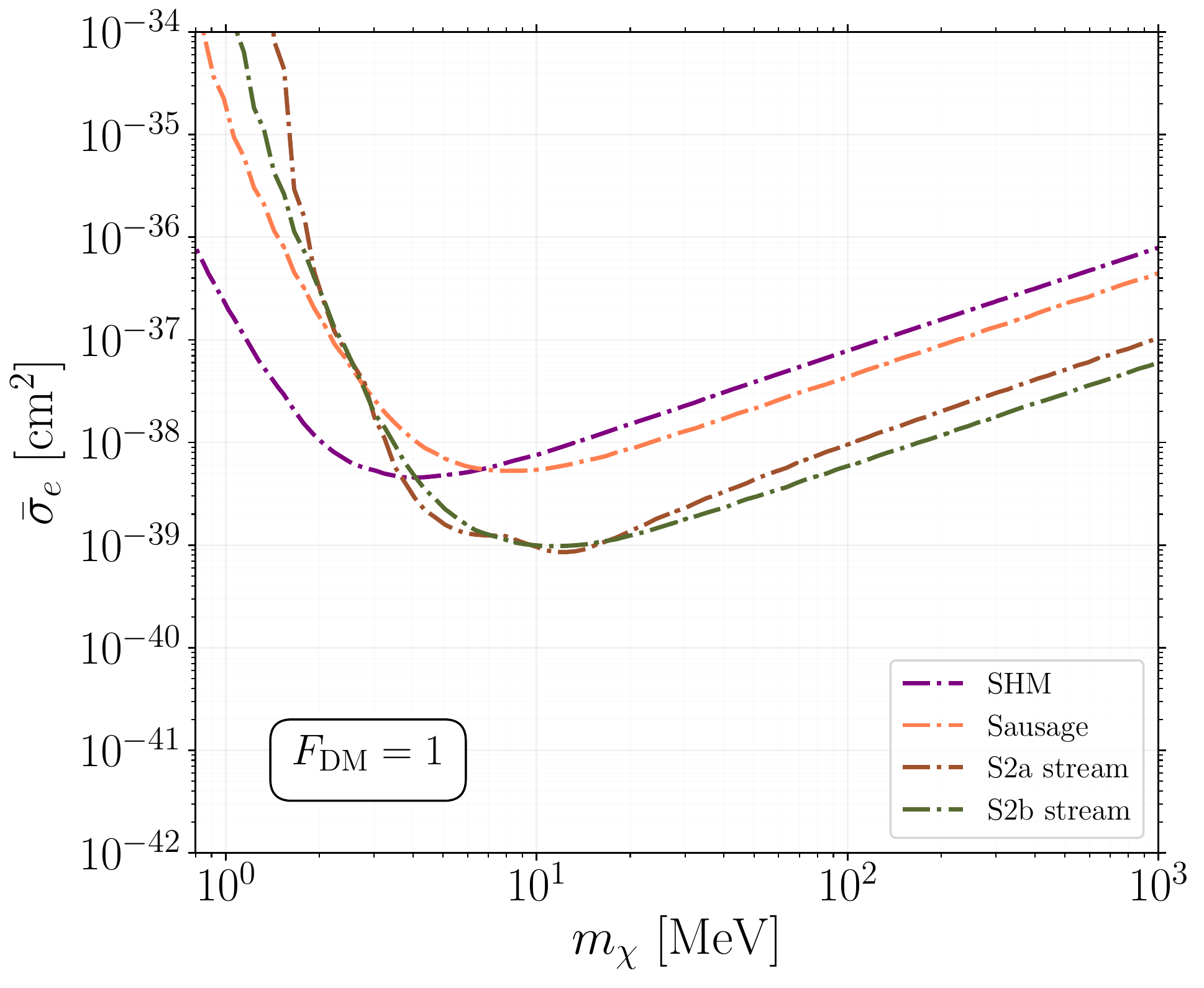}
    \caption{Discovery reaches for DM (non-modulating) signal and modulation at a SENSEI-like experiment with 1 kg-year exposure shown in the DM mass -- cross section plane. {\it Top row:} Dashed (solid) yellow contours show the 5$\sigma$ discovery reach using a MAX (OPT) background model, whereas the modulation discovery is indicated by the dot-dashed purple contour; these are plotted for DM form factors, ({\it left}) $F_{\rm DM}= 1$ and ({\it right}) $F_{\rm DM} \propto 1/q^2$, assuming the SHM velocity distribution. Also shown for reference are the latest 90$\%$ CL upper limits from SENSEI~\cite{Barak:2020fql} (cyan), XENON10~\cite{Essig:2017kqs} (orange), and XENON1T~\cite{Aprile:2019xxb} (gray).\footnote{ Despite the more conservative bound shown by XENON1T in Fig. 5 of Ref.~\cite{Aprile:2019xxb}, we only show the constraints for events with $\geq 12 \, e^-$, in case of $F_{\rm DM}= 1$, because the charge yield for liquid xenon has never been measured below these energies. Meanwhile, for $F_{\rm DM} \propto 1/q^2$, the corresponding constraint of XENON1T lies above the cross section range of the plot.} {\it Bottom row:} Modulation discovery reaches for: ({\it left}) various fractions of Nyx and S1 streams represented by blue and red contours; ({\it right}) S2a and S2b streams along with the {\it Gaia} Sausage indicated by brown, green, and orange contours respectively. All the contours for DM substructure components assume that it constitutes 100\% of the local DM density.}
    \label{fig:dr_compare}
\end{figure*}

From a statistical point of view, our Asimov discovery reaches indicate the median sensitivity of a SENSEI-like experiment to a DM signal. Before discussing the case of different DM substructure components, however, we study the basic characteristics of discovery contours for the vanilla SHM velocity distribution.

In the top row of Fig.~\ref{fig:dr_compare}, we plot two types of 5$\sigma$ discovery reach contours for different DM form factors: one for detecting DM over a given background model (MAX or OPT), and the other for observing a modulating DM signal over the average recoil spectrum. Since the modulation fraction for DM-$e$ scattering is $\mathcal{O}(1$-$10)\%$, we would naively expect the modulation sensitivity to be at least an order of magnitude weaker, {\it i.e}, it requires a larger cross section to achieve a similar significance of detection, than that of simply discovering DM. The presence of experimental backgrounds, however, leads to two important features. First, the DM discovery reach using the MAX model is significantly weaker than the one using the OPT model for $m_\chi \gsim 3$ MeV. Second, for lower masses, $m_\chi \lesssim 5$ MeV, an experiment is more likely to discover a modulation signal first. To explain these behaviors, we note that the recoil spectra for DM-$e$ scattering peaks in the first couple of $Q$ bins for masses $m_\chi \lesssim \mathcal{O}(10)$ MeV, with the peak shifting to higher $Q$ bins for increasing DM masses where the background rate is dominated by the constant rate component. At larger DM masses, the TS of discovery reach roughly has the usual $\sqrt{B}$ scaling. Since the constant rate in the MAX model is $\sim 10^3$ greater than in the OPT model, we can see that the discovery reach for a general DM signal is worse by a factor of 30 in the MAX model than that in the OPT model. Meanwhile, at lower masses where $Q= 1, 2$ bins drive the discovery reach, the extremely high background rate in the MAX model and the higher threshold of the OPT model lead to considerably weaker reaches for a non-modulating signal for both background models, compared to the modulating one. The discovery contours follow a similar trend for $F_{\rm DM} \propto 1/q^2$, except that they are marginally stronger (weaker) at lower (higher) DM masses. Again, this is explained by the `squeezing' of the recoil spectrum (more events in the peak, fewer in the tail) to lower $Q$ bins with an enhanced crystal form factor at low $q$, due to the DM form factor.

Next, we investigate how the unique kinematic features of various DM substructure components discussed in Sec.~\ref{subsec:astro_spectrum} will affect their detection prospects in the bottom row of Fig.~\ref{fig:dr_compare}. We focus on the discovery of a modulation signal since it can effectively be considered as a background-free channel. The most striking feature for all substructure components is the marked deviation of their modulation discovery reach compared to the SHM contour.\footnote{However, as shown by Fig.~\ref{fig:dr_compare_app} in Appendix~\ref{appD}, that is not the case for DM discovery reach of substructure components with lower $\vmp$ than SHM for both MAX and OPT background models.} For example, relative to SHM, the reaches for Nyx, S2a, S2b, and Sausage, all weaken (strengthen) for low (high) DM masses with an inflection point at $m_\chi \sim 10$ MeV. Upon closer scrutiny, we observe that Nyx, S2a, S2b have similar sensitivity at high DM masses whereas the Sausage contour is quite a bit weaker, while at the low mass end, amongst them the reaches follow the order (in increasing strength): S2a $<$ Nyx $<$ S2b $<$ Sausage. These behaviors could be understood as follows:\footnote{We also tested our explanations on the discovery reaches for several simulated toy streams with artificial velocity dispersions and phases, and found that it is able to satisfactorily explain the qualitative differences between their sensitivities.} {\it i)} the amplitude of modulation approximated by Eq.~\eqref{ampl}, which depends on $\vmp$ and the coplanarity $b$, is lower for SHM and Sausage relative to the streams (other than S1), leading to weakened sensitivities to these astrophysical components at higher mass; {\it ii)} since $v_{\rm min}$ at low DM masses lies in the tail of the velocity distribution, substructure components with greater $\vmp$ and/or velocity dispersion $\sigma_v$ (see Table~\ref{table:distr_pars}) have a stronger discovery reach, explaining the order of reaches at the low mass end. We can also apply these heuristics to understand qualitatively the behavior of the S1 stream as well. It has an increased sensitivity at low DM masses due to its high $\vmp$ with almost an order of magnitude improvement over SHM at $m_\chi \sim 2$ MeV. But when combined with its small coplanarity $b$, the suppressed amplitude of modulation leads to a much weaker reach at high DM masses, compared to the other streams.

A major caveat for the discussion above is the implicit assumption that DM in each component constitutes 100$\%$ of the local DM density. We also show in the bottom left panel the discovery reaches for an astrophysical model where only 20$\%$ of the DM lies within Nyx or S1 streams and SHM contributes the remaining fraction. Although there appears to be very little difference in these discovery contours, as we argue in the following sections, DM-$e$ recoil experiments could still play a significant role in constraining the DM fraction in streams if we assume the discovery of a signal.

\subsection{Resolving DM substructure fraction}
\begin{figure*}[tbh]
    \centering
       	 \includegraphics[width=0.48\textwidth]{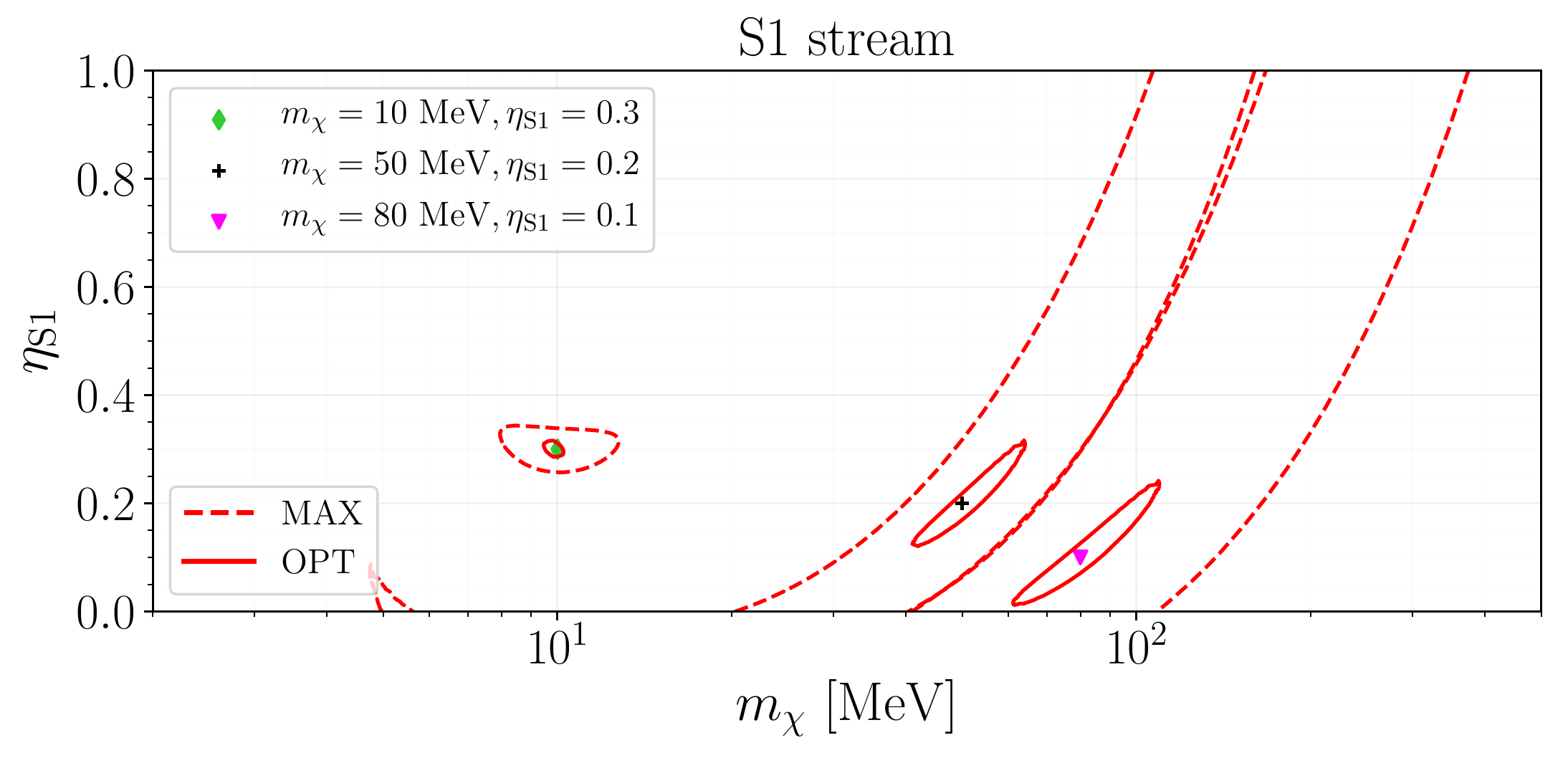}
         \includegraphics[width=0.48\textwidth]{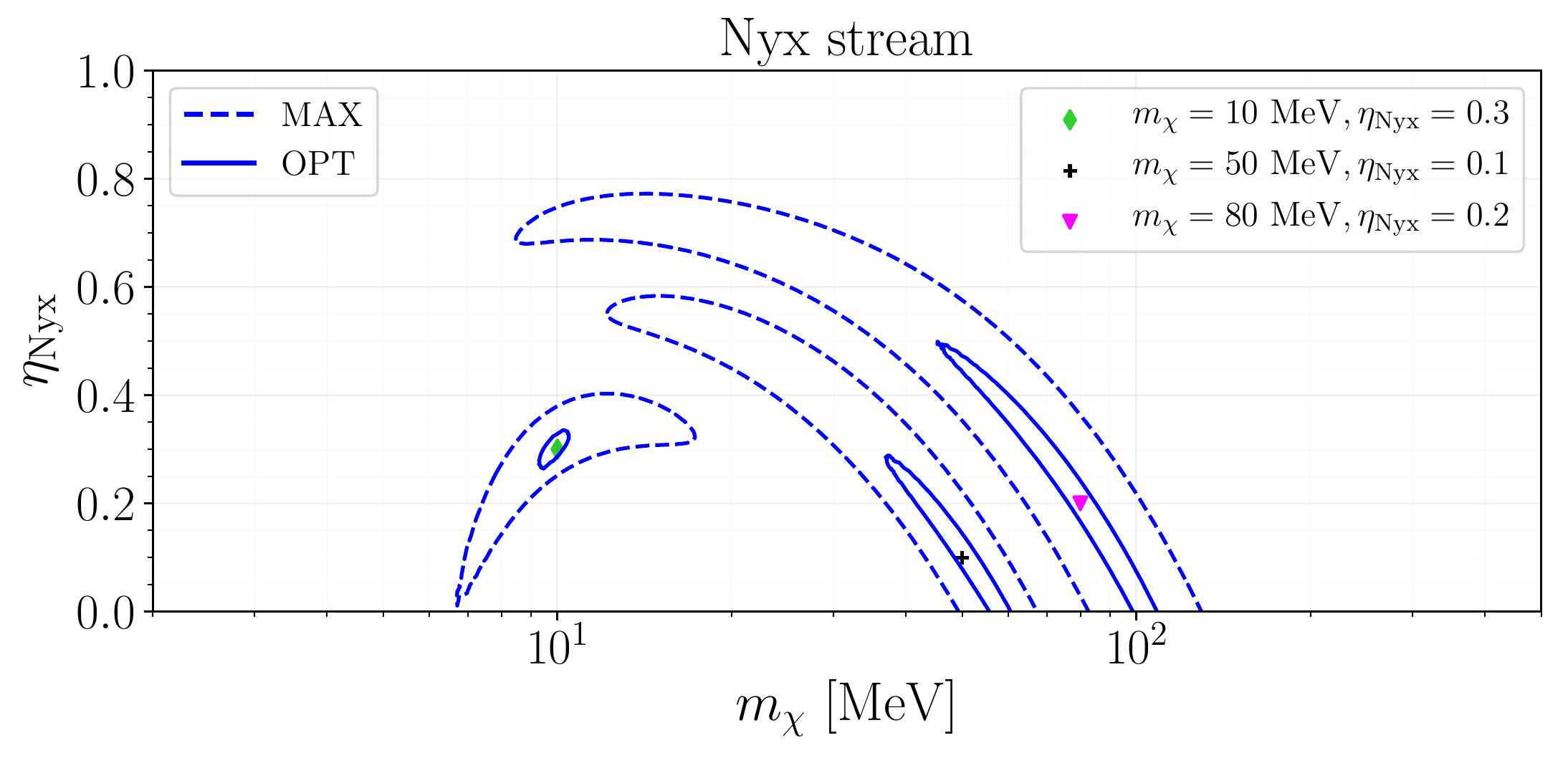}
    	\includegraphics[width=0.48\textwidth]{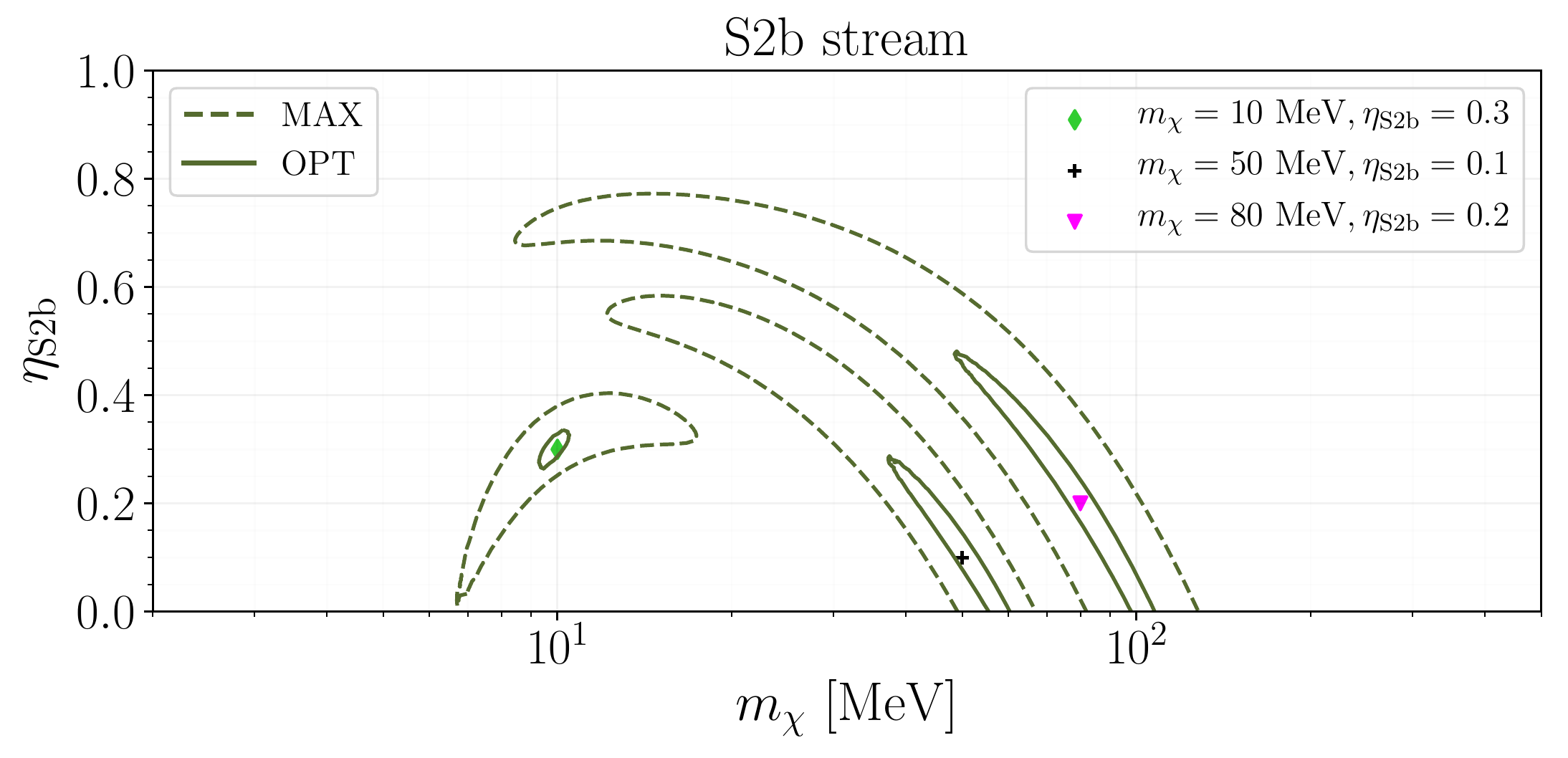}
    	 \includegraphics[width=0.48\textwidth]{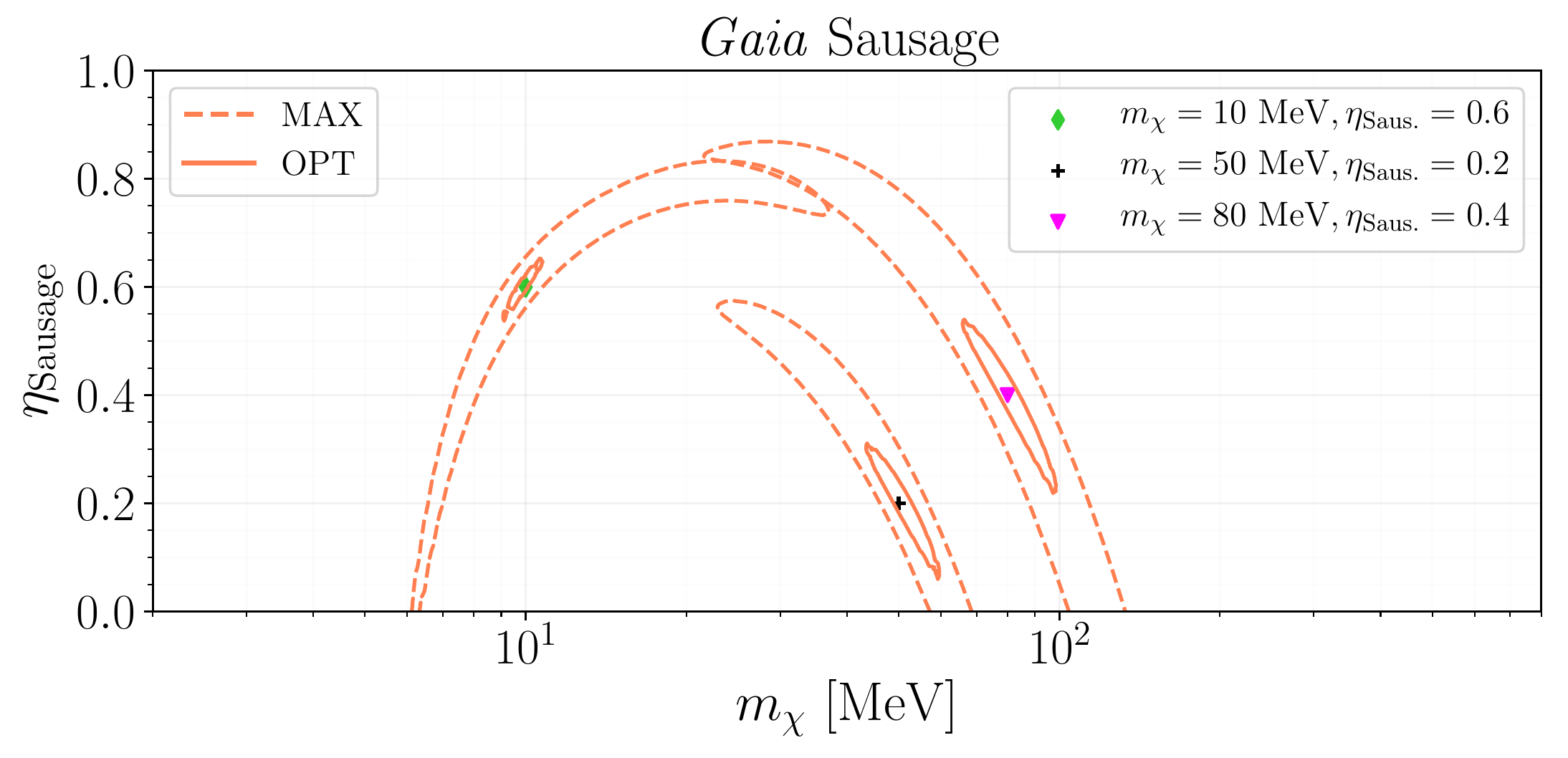}
    \caption{68$\%$ CL contours in the DM mass -- DM substructure fraction plane at different benchmark points for S1 ({\it top left}), Nyx ({\it top right}), S2b ({\it bottom left}) streams, and {\it Gaia} Sausage ({\it bottom right}). The remaining $(1 -\eta_i)$ fraction of DM is constituted by the SHM and the smooth isotropic halo for streams and Sausage respectively. As in Fig.~\ref{fig:dr_compare}, the solid (dashed) contours denote the forecasts for the OPT (MAX) background model. The benchmark scattering cross section for all the substructure components is $\bar{\sigma}_e = 10^{-38} \, {\rm cm}^2$, which is in the neighborhood of the 5$\sigma$ modulation discovery reach for the benchmark masses considered here.}
    \label{fig:contours_em}
\end{figure*}

Using the average DM-$e$ recoil spectrum in Eq.~\eqref{eq:mod_compare}, we forecast the sensitivity of a next-generation SENSEI-like experiment to reconstruct the DM fraction $\eta_i$ in the $i^{\rm th}$ DM substructure component. In particular, we consider how the resolution of $\eta$ varies with DM mass in the case which contains only a single DM substructure component and a total velocity distribution given by
\beq
\eta f_{\rm sub} (\vec{v}) + (1- \eta) f_{\rm SHM} (\vec{v}).
\eeq
In addition, we try to simultaneously constrain $\eta$'s for components in a dual substructure model with the velocity distribution given by
\beq
\sum_{i=1}^2\eta_i f_{i} (\vec{v}) + (1- \eta_1-\eta_2) f_{\rm SHM} (\vec{v}),
\eeq
where the subscript $i= 1, 2$ refers to any two distinct DM substructure components. The real astrophysical composition of the local DM distribution could be more complicated, and the two cases we consider just serve as illustrative examples of how DM-$e$ scattering experiments could probe DM substructures. For concreteness, we only discuss the results for $F_{\rm DM} = 1$.

In Fig.~\ref{fig:contours_em}, we show the 68$\%$ CL sensitivity forecast contours using MAX and OPT background models for S1, Nyx, S2b, and Sausage substructure components. 
We find that the next generation SENSEI-like experiment with 1 kg-year exposure can narrow down DM substructure fractions as a function of $m_\chi$. In particular if we assume an optimistic background, we can localize $\eta$ given a perfect knowledge of the DM substructure distribution. In a more realistic interpretation, when our knowledge of the DM velocity distribution is not perfect, we still expect a reasonable resolution of $\eta$ at $m_\chi \lesssim 20$ MeV, with either background model. Our result should still hold qualitatively in the situation when the velocity distribution of a dark stream does not perfectly correlate with the stellar one.

We also comment upon several general features of the degeneracy contours in Fig.~\ref{fig:contours_em}. First, the experimental sensitivity progressively worsens in both $m_\chi$ and $\eta_i$ directions when increasing the DM mass for all four components, whereas the resolution is fairly good for all streams at low DM masses even with the MAX background model. Broadly speaking, these features can be interpreted in terms of the DM mass and velocity dependence of the recoil rate. The overall scattering rate could be kept roughly similar when varying DM mass and substructure fraction simultaneously. Yet at larger DM masses above 20 MeV, spectra shapes are degenerate for a fairly large range of masses and cannot be fully broken by varying the substructure fraction. While for $m_\chi \lesssim 20$ MeV, the spectral shape in high $Q$ bins is quite sensitive to $\eta$, which strongly affects the number of DM particles in the tail of the velocity distribution as the substructure could have a quite different value of $\vmp$ with respect to the remaining SHM/halo component.

One may also notice the intriguing differences between the orientations of contours at the benchmark points with the same DM masses for S1 on one hand and for Nyx, S2b, and Sausage on the other hand. For the latter class of DM components, for a DM mass of 10 MeV, our sensitivity forecasts indicate a degeneracy between $\eta$ and $m_\chi$ with a preference for higher DM masses at higher DM substructure fractions, while the forecast for S1 is largely independent of $\eta$. At small masses around and below 10 MeV, $\vmin$ is large for given $(q, E)$ and thus it only probes the tail of $g(\vmin)$. Since Nyx, S2b, and Sausage all have a lower $\vmp$ compared to that of SHM, increasing their $\eta$ at a fixed $m_\chi$ leads to a smaller tail for $g(\vmin)$ of the combined velocity distribution. This effect is compensated by lowering $v_{\rm min}$ and enhancing $g(\vmin)$ when increasing $m_\chi$. Conversely, S1's high $\vmp$ implies a much wider plateau as shown in Sec.~\ref{subsubsec:average}, which allows for more scattering at low DM masses leading to a considerably better resolution. Note that for higher DM masses, the orientation of the contours is flipped, {\it i.e} when increasing $\eta$, the S1 contours tend toward higher values of $m_\chi$, whereas those for Nyx, S2b, and Sausage have the opposite behavior. At larger DM masses, where $v_{\rm min}$ is sufficiently small, it is the height of the plateau in $g(\vmin)$ that dictates the orientation. As we note from Fig.~\ref{fig:f2_gvmin}, components with lower $\vmp$ such as Nyx have a higher plateau relative to those with a higher $\vmp$ such as SHM and S1. Thus, the rate will increase with an increase in $\eta$ for Nyx, S2b and Sausage at a given large DM mass. This effect could be cancelled by increasing $\vmin$ through decreasing $m_\chi$. Opposite arguments apply to the S1 stream.

Lastly, as was highlighted previously, the S1 stream peaks at higher $Q$ bins relative to components with lower values of $\vmp$. This creates a significant improvement in its resolution, especially at higher DM masses, when switching from the MAX to the OPT background model. There is also an improvement for other substructure components, but it is less striking.

\begin{figure*}[tbh]
    \centering
       \includegraphics[width=0.48\textwidth]{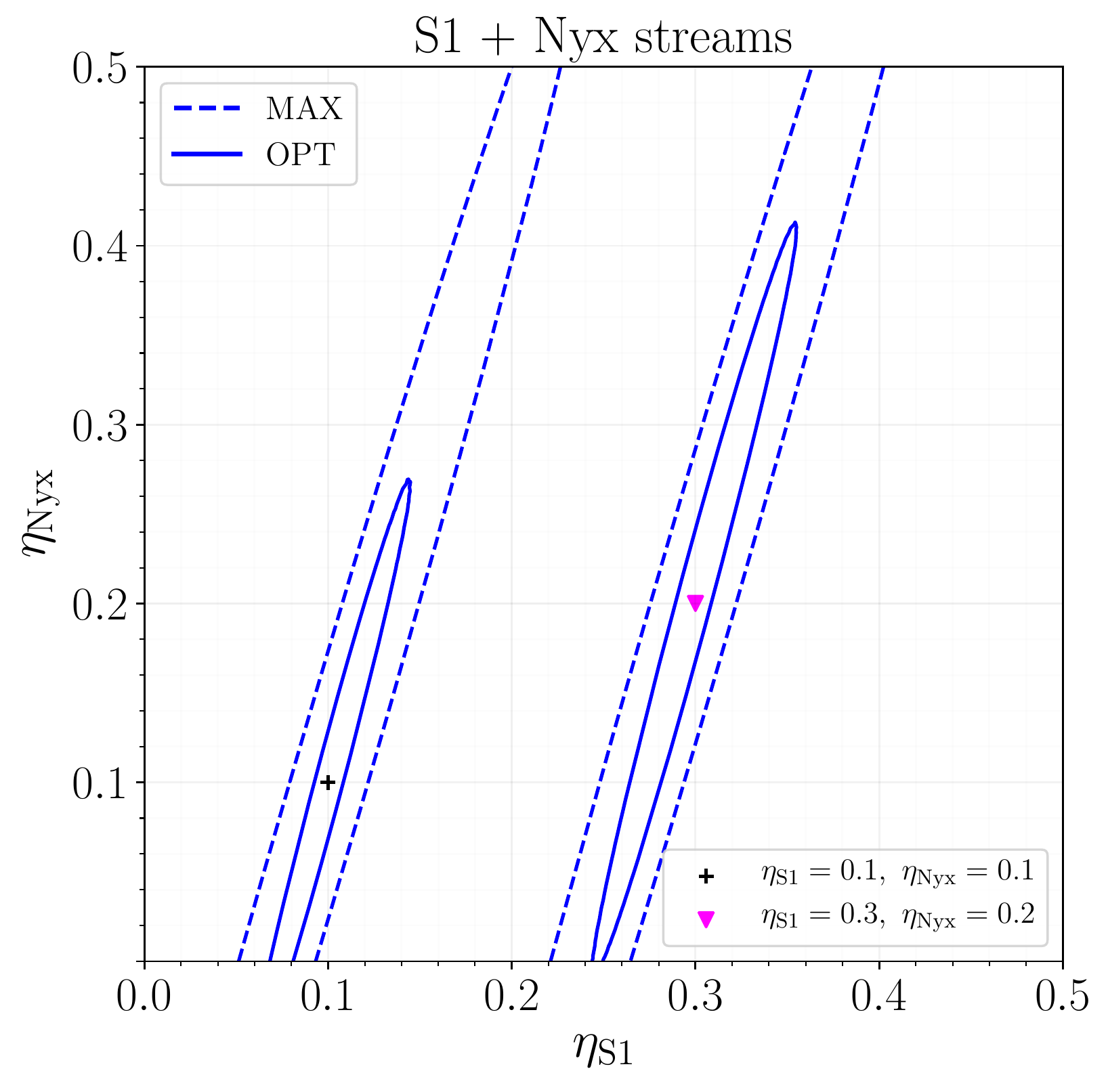}
    	\includegraphics[width=0.48\textwidth]{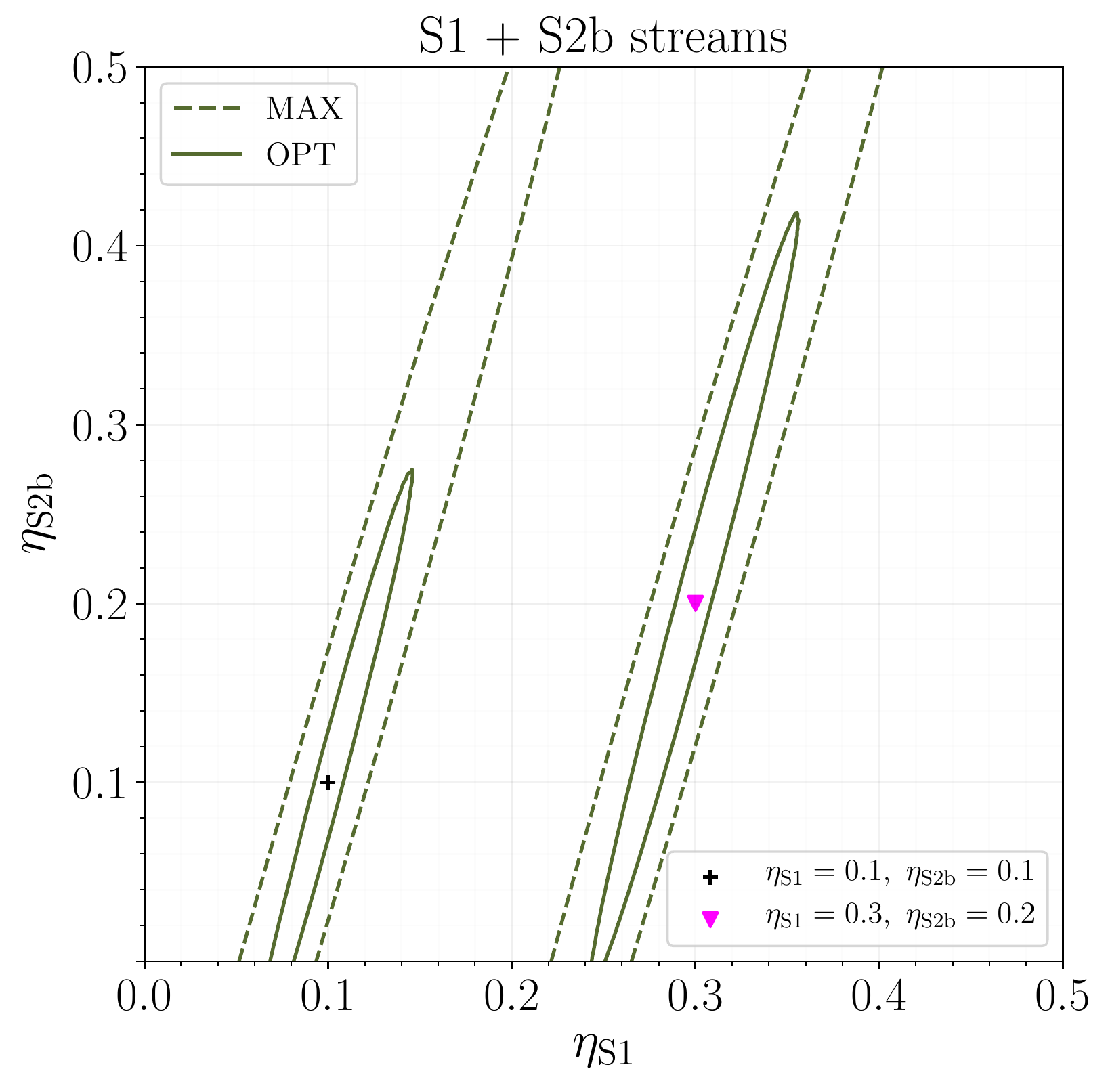}
   	    \includegraphics[width=0.48\textwidth]{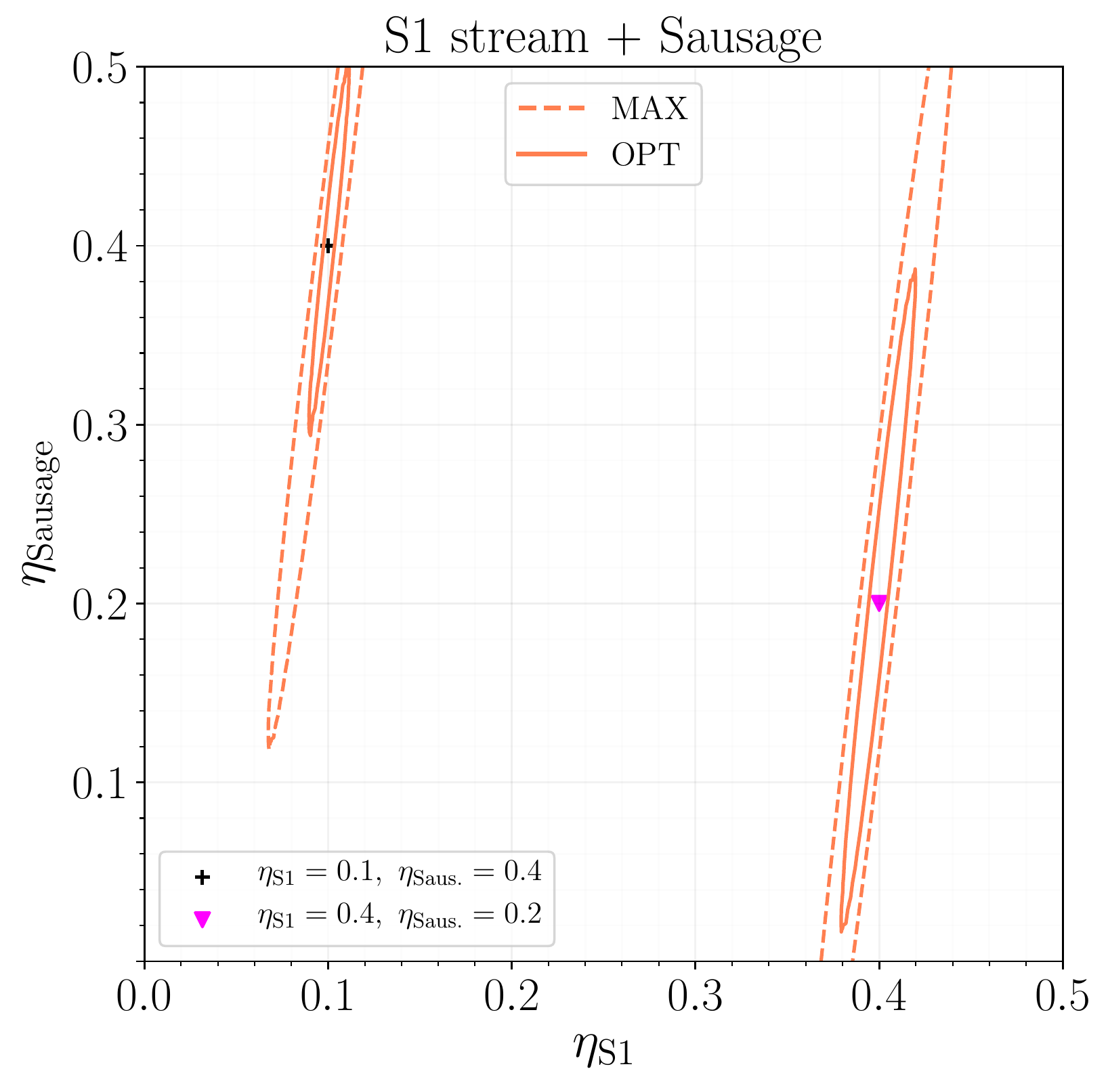}
   	     \includegraphics[width=0.48\textwidth]{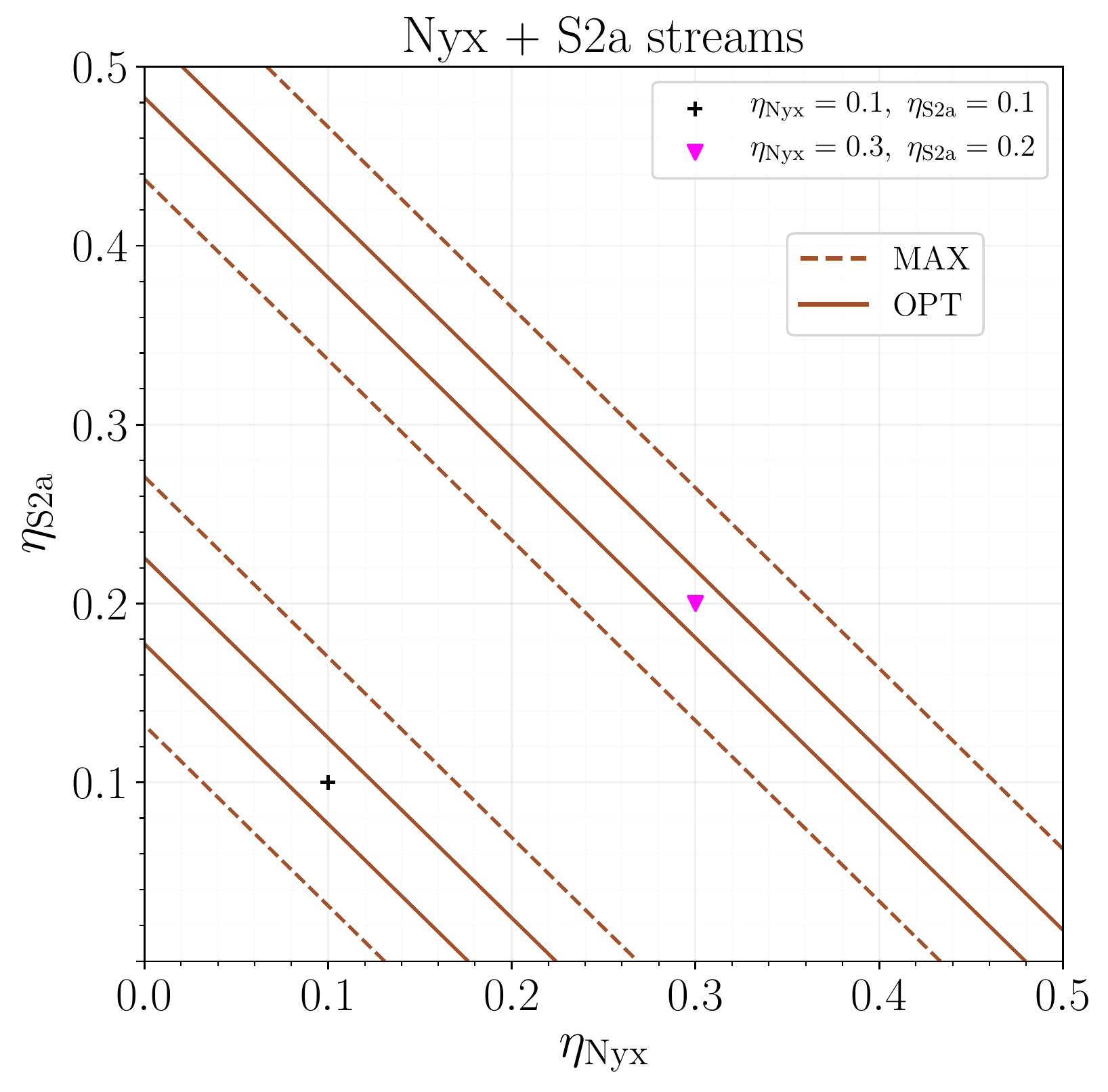}
    \caption{68$\%$ CL contours for resolving DM fractions in astrophysical configurations with two substructure components: S1 and Nyx streams ({\it top left}), S1 and S2b streams ({\it top right}), S1 stream and {\it Gaia} Sausage ({\it bottom left}), and Nyx and S2a streams ({\it bottom right}). The remaining $(1 -\eta_i)$ fraction of DM is constituted by the SHM for all stream combinations considered here, and by the smooth, isotropic halo for the combination of S1 stream and Sausage. As in Fig.~\ref{fig:dr_compare}, the solid (dashed) contours denote the forecasts for the OPT (MAX) background model. The benchmark mass and scattering cross section for all the plots are $m_\chi = 20$ MeV and $\bar{\sigma}_e = 10^{-38} \, {\rm cm}^2$ respectively.}
    \label{fig:contours_ee}
\end{figure*}

So far our discussion has focused only on the toy scenario where the local DM astrophysical distribution consists of just a smooth halo along with one additional substructure component. Next, we explore the possibility of simultaneously constraining the substructure fractions of a local DM distribution with two additional components. Fig.~\ref{fig:contours_ee} shows the results of our analysis through 68$\%$ CL sensitivity forecasts in $\eta$--$\eta$ parameter space for $m_\chi =20$ MeV. As we can see from Fig.~\ref{fig:contours_em}, the electron recoil experiment has maximum sensitivity to the DM substructure fraction in the neighborhood of $m_\chi =20$ MeV, implying that our forecasts in Fig.~\ref{fig:contours_ee} are on the optimistic side. For three of our plots, we have fixed S1 as one of the components, combining it with Nyx, S2b, or Sausage. Since all components besides S1 have similar values of $\vmp$, their degeneracy contours have similar characteristics. In particular, we note that the contours are very sensitive to changes in $\eta_{\rm S1}$, while being highly degenerate in the fraction of the other component for the MAX background model. Moreover, there is a moderate improvement in the resolution when we use the OPT background. The sensitivity to $\eta_{\rm S1}$ can be clearly attributed to S1's distinct $\vmp$, which is the highest among all components for any configuration. In the remaining plot between Nyx and S2a, we see that the substructure fractions are maximally degenerate for both background models, given the similar $\vmp$ values of both streams. This degeneracy for $m_\chi \gsim 5$ MeV could be mitigated by reducing the background in $Q=1, 2$ bins, despite the difference in their $\vmp$ values being only $\approx 80$ km/s.

While we take the mean values of stellar substructures as benchmarks of their DM counterparts, we do not include their uncertainties in our analysis. The magnitudes of these uncertainties are provided in Table~\ref{table:distr_pars}. Incorporating them will not change our results much. As we have repeatedly emphasized, potentially the most important (systematic) uncertainty is the correlation between stellar and DM streams. We leave a more quantitative uncertainty analysis for the future, when the DM-stellar correlation in streams is better understood.

Finally, we comment on the cross section, $\bar{\sigma}_e = 10^{-38} \, {\rm cm}^2$, used for obtaining the results discussed in this section. Since our benchmark cross section is in the neighborhood of the 5$\sigma$ modulation discovery reach for all the benchmark masses we consider in Figs.~\ref{fig:contours_em} and ~\ref{fig:contours_ee}, it is an intriguing prospect to see whether adding time domain information leads to better signal discrimination, or equivalently an improvement in the resolution of $\eta$. However, we have verified that for the different substructure combinations considered here, the effect of $\vmp$ always dominates those of $t_c$ and $b$, even when we assume a negligible background in the $Q= 1, 2$ bins. An alternative way to extract maximum information from time domain, particularly while using actual experimental data, is to consider correlations between time bins since these could be unique for each DM substructure component irrespective of their $\vmp$ value. Such a study is beyond the scope of our forecasting analysis and we defer it to a future work.

\section{Conclusions and Outlook}
\label{sec:discussion}

The rich galactic dynamics of phase space substructure revealed by {\it Gaia} data suggests a potentially more complicated composition of DM around us beyond the simple description of SHM. This could have important implications for terrestrial DM probes, e.g., DM-$e$ scattering experiment, which is the low-mass frontier of DM direct detection.

In this article, we first study how possible new substructure components could affect the observables of DM-$e$ scattering, both the time average recoil spectra and the yearly modulation. One could understand these effects through three quantities that characterize the substructure components: most probable speed $\vmp$, characteristic time $t_c$ and coplanarity $b$. We then perform a likelihood-based analysis to demonstrate how the discovery reach of a future DM-$e$ scattering experiment depends on different astrophysical DM models (see also a complementary study~\cite{yu2020}). In particular, we show that given a discovery, DM-$e$ scattering experiments could be sensitive to one or several DM substructure components, and will be able to constrain their corresponding fractions even when they are sub-dominant to the local DM density. This suggests an interesting opportunity to probe the astrophysical aspects of DM models using direct detection experiments.

The relationship between direct detection and local DM distributions is still an evolving subject -- one that requires further study. In our study, we take the data of the stellar streams, identified using {\it Gaia} data, as proxies of their possible DM counterparts, and explore their potential effects at future DM-$e$ scattering experiments. Yet the correlation between the stellar and associated DM substructure is not fully established. It is important to test and confirm the properties of DM substructure with further observations and numerical simulations.

Given our benchmarks, we find that DM-$e$ experiments could probe and constrain fractions of DM in substructure, driven largely by their different $\vmp$'s. It will be interesting to devise more sophisticated statistical analysis to take advantage of effects due to $t_c$ and $b$. Ideally, it will be fantastic to apply these methods on actual experimental data to probe the local DM distribution.

\textbf{Acknowledgements} We thank Lina Necib and Ciaran O'Hare for useful correspondence and for sharing their velocity distributions data files. We are also grateful to Tien-Tien Yu for offering critical comments on a preliminary draft of the manuscript. The results in this work were computed using the following open-source software: \texttt{swordfish}~\cite{Edwards:2017kqw}, \texttt{IPython}~\cite{Perez:2007emg}, \texttt{matplotlib}~\cite{Hunter:2007ouj}, \texttt{scipy}~\cite{jones2001scipy}, and \texttt{numpy}~\cite{vanderWalt:2011bqk}. JB, JF, JL, and M. B-A are supported by the DOE grant DE-SC-0010010 and NASA grant 80NSSC18K1010.

\appendix
\section{Astrophysical components}
\label{appA}
In this paper, we consider several possible astrophysical components in the solar neighborhood that have been discussed in the literature. We include the \textit{Gaia} Sausage for tidal debris, a kinematic substructure resulting from older mergers which becomes well-mixed spatially and only manifests itself in the velocity distribution~\cite{Lisanti:2011as}. A distinctive feature of the \textit{Gaia} Sausage is that there are two lobes in the radial velocity distribution, at $v_r = \pm 115.50$ km/s.

We also consider three stellar streams, which are kinematically cold substructures that are localized in both position and velocity space: \textit{i)} Nyx, a prograde stream with $\sim$500 stars that slightly lags behind the MW disk~\cite{Necib:2019zbk}; \textit{ii)} S1, a retrograde stream with 28 stars and a very high Earth-frame speed~\cite{OHare:2018trr}; and \textit{iii)} S2, a stream following a prograde orbit with a high vertical direction component. S2 has two constituents, S2a with 48 stars and S2b with 8 stars. S1 and S2 streams are two of the most prominent streams belonging to a group of substructures referred to as stellar shards~\cite{OHare:2019qxc}.

Lastly, we include the halo component of the distributions. We use the SHM parameterized as in Ref.~\cite{Freese:2012xd}. We also use, in conjunction with \textit{Gaia} Sausage, the \textit{Gaia} halo described in Ref.~\cite{Necib:2018iwb}, which comes from the joint posterior for the halo component of the MW when modeling the \textit{Gaia} Sausage substructure. Note that \textit{Gaia} halo is different from SHM.

The full parameters of the velocity distributions of these components (in the galactic frame) are listed in Table~\ref{table:distr_pars}. The mean values for the parameters with an asterisk are obtained by fitting Gaussian distributions to the kinematic data provided by Lina Necib, from Refs.~\cite{Necib:2018iwb,Necib:2019zbk}. Also included in this table are the uncertainties for all the parameters. The uncertainties in the parameters with an asterisk were derived by propagating the errors in the published literature to our fitted values \cite{Necib:2018iwb,Necib:2019zbk}; while the uncertainties in the shards streams were provided by Ciaran O'Hare, from Ref.~\cite{OHare:2019qxc}.

Since both tidal debris and streams are remnants of accretion from MW's surrounding satellites and subhalos, we expect that there are DM counterparts to the stellar components. The correlation between the velocity distributions of stellar components and their DM counterparts have been discussed for some substructures such as \textit{Gaia} Sausage~\cite{Bozorgnia:2018pfa,Lisanti:2014dva,Necib:2018igl} and requires further investigation. While simulation shows that debris flow could be a good tracer of DM~\cite{Lisanti:2014dva,Necib:2018igl}, the relation between DM and stellar components of streams is {\it not} established yet. Nonetheless, we use velocity distributions of stellar streams as tentative descriptions of their associated DM component. The purpose is to use these as benchmarks to show that DM substructure could have interesting distinctive effects on direct detection experiments while on the other hand, terrestrial DM experiments could probe astrophysical DM substructures. For a more precise description of DM substructure (especially DM streams), we need to wait for numerical simulations in the near future.

\begin{table*}[!ht]
\centering
\def\arraystretch{1.25}
\begin{tabular}{ | c  | c  | c | }
\hline
& Mean Velocity & Velocity Dispersion  \\ [-3pt]
Component & $(\mu_r, \mu_\phi, \mu_z) \pm (\Delta\mu_r, \Delta\mu_\phi, \Delta\mu_z)$ & $\mathrm{diag}(\sigma_r, \sigma_\phi, \sigma_z) \pm (\Delta\sigma_r, \Delta\sigma_\phi, \Delta\sigma_z)$  \\  [-2pt]
 & $[\km/\seg]$ &  $[\km/\seg]$  \\ [3pt] \hline
SHM~\cite{Freese:2012xd} & $(0, 0, 0)$ & $(155.6, 155.6, 155.6) \pm (2.12, 2.12, 2.12)$  \\ [0.5pt]
GAIA halo~\cite{Necib:2018iwb}* & $(0, 0, 0)$ & $(143.96, 132.03, 118.30)^{+(4.31, 4.30, 3.42)}_{-(5.03, 3.57, 1.86)}$  \\ [0.5pt]
GAIA sausage~\cite{Necib:2018iwb}* & $(\pm 115.50, 36.94, -2.92)^{+(1.77, 1.87, 0.85)}_{-(2.06, 1.87, 0.85)}$ & $(108.33, 62.60, 57.99)^{+(1.20, 1.53, 0.70)}_{-(1.30, 1.53, 0.80)}$  \\ [0.5pt]
Nyx~\cite{Necib:2019zbk}* & $(133.90, 130.17, 53.65)^{+(1.79, 2.31, 118.79)}_{-(1.88, 2.40, 114.96)}$ & $(67.13, 45.80, 65.82)^{+(2.43, 1.57, 2.23)}_{-(2.29, 1.57, 2.04)}$  \\ [0.5pt]
S1~\cite{OHare:2019qxc}  & $(-34.2, -306.3, -64.4) \pm ( 27.92, 21.34, 18.34)$ & $(81.9, 46.3, 62.9) \pm (22.76, 32.37, 23.35)$\\ [0.5pt]
S2a~\cite{OHare:2019qxc}   & $(5.8, 163.6, -250.4) \pm ( 18.34, 18.52, 19.84) $ & $(45.9, 13.8, 26.8) \pm (17.13, 12.86, 15.66)$ \\ [0.5pt]
S2b~\cite{OHare:2019qxc}  & $(-50.6, 138.5, 183.1) \pm (16.33, 15.21, 20.16)$ & $(90.8, 25.0, 43.8) \pm (21.77, 12.05, 13.66)$ \\ \hline
\end{tabular}
\caption{Parameters describing velocity distributions of different possible DM components in the solar neighborhood, in the galactic rest frame. The asterisk data are extracted from Gaussian-fits to the star data provided by Necib et al.\footnote{Note that the mean values are different from the values quoted in Refs.~\cite{Necib:2018iwb,Necib:2019zbk}.} The SHM velocity dispersion uncertainty is derived from Ref.~\cite{2017MNRAS.465...76M}.}
\label{table:distr_pars}
\end{table*}

Based on the discussion above, we now write the velocity distribution $f_i(v)$ of the $i$-th astrophysical component in the Earth's (lab's) frame:
\begin{widetext}
\beq\label{lab_distr}
	f_i(\vec{v}) = \frac{1}{N_{i, {\rm esc}}} \frac{1}{\sqrt{(2\pi)^3 \det \cov_i}} \exp \bl[ -\frac{1}{2}(\vec{v} - \vec{\mu}_{i, {\rm lab}}(t) )^T \cdot \cov_i^{-1} \cdot (\vec{v} - \vec{\mu}_{i, {\rm lab}}(t) ) \br] \Theta(\vesc - \vert\vec{v} + \vec{v}_{\rm lab}(t) \vert) \ ,
\eeq
\end{widetext}
where $\cov_i \equiv \mathrm{diag}(\sigma_r^2, \sigma_\phi^2, \sigma_z^2)$ is the square of the velocity dispersion matrix, and $\vec{\mu}_{i, {\rm lab}}(t)$ is the mean velocity of the DM wind boosted to the lab frame:
\beqa\label{mu_lab}
  \vec{\mu}_{i, {\rm lab}}(t) & \equiv & \vec{\mu}_i - \vec{v}_{\rm lab}(t) \ , \\
  \quad \text{with} \quad \vec{v}_{\rm lab}(t) & \equiv & \vec{v}_\Sun + \vec{V}_\Earth(t) \ .
\eeqa
$\vec{v}_\Sun$ is the Sun's velocity in the galactic rest frame, and $\vec{V}_\Earth(t)$ is the Earth's velocity in the heliocentric frame. Following \cite{2010MNRAS.403.1829S, 2017MNRAS.465...76M, 2019A&A...625L..10G}, we take these to be:
\beqa\label{vsun}
    \vec{v}_\Sun & = & (U, V, W)  =  (11.1, 247.24, 7.25)~\km/\seg \ , \\
    \vec{V}_\Earth(t) & = & V_\Earth \left[ \vec{\epsilon}_1 \cos(\omega(t-t_{\rm Mar~21})) \right. \nn\\
    & & \left. + \vec{\epsilon}_2 \sin(\omega(t-t_{\rm Mar~21})) \right] \ ,\label{vearth}
\eeqa
with $\omega = 2\pi/365.25~{\rm days}^{-1}$ the Earth's angular speed, and $V_\Earth=29.79~\km/\seg$ its orbital speed. $\vec{\epsilon}_{1,2}$ are two linearly independent vectors defining the Earth's circular orbit (ignoring its eccentricity) which, in the conventions of Refs.~\cite{Green:2003yh, Lee:2013xxa}, point in the direction of the Earth's velocity during the vernal equinox ($t_{\rm Mar~21} = 79.26~{\rm days}$) and the summer solstice, respectively:
\beqa\label{epsilons}
  \vec{\epsilon}_1 & = & (0.9940, 0.1095, 0.0031) \ , \\
  \vec{\epsilon}_2 & = & (-0.0517, 0.4945, -0.8677).
\eeqa
Finally, $\Theta$ in Eq.~\eqref{lab_distr} is a Heaviside step function that cuts off the DM speed at escape velocity $\vesc$, which throughout this paper we take to be $528~\km/\seg$ \cite{2019MNRAS.485.3514D}. Due to this velocity cut-off, we need a constant normalization factor $N_{i, {\rm esc}}$ for each DM component.

\section{Characteristic quantities}
\label{appB}

In section~\ref{subsec:astro_setup}, we introduced the most probable speed $v_{\rm mp}$, characteristic time $t_c$, and coplanarity $b$ to describe all the DM components we have considered. We present their precise definitions below.

The definition of $v_{\rm mp}$ is
\beqa\label{vmp}
v_{\rm mp} & \equiv & \underset{v}{\operatorname{arg\,max}}\,(\overline{F}(v)) \nonumber \\
&=&  \underset{v}{\operatorname{arg\,max}}\,\frac{1}{\rm{year}}\!\int\limits^{\rm{year}}_0\!\!\mathrm{d}t\,F(v,t).
\eeqa

For a DM component with a mean velocity $\vec{\mu}_{\rm lab}(t)$ in the Earth's frame, given by Eq.~\eqref{mu_lab}, the characteristic time is
\beq\label{char_time}
t_c \equiv \underset{t}{\operatorname*{arg\,max}}\, (\vec{\mu}_{\rm lab}(t)).
\eeq
The coplanarity $b$ is a measure of whether the DM wind lies in the same plane of the Earth's orbit. It is given by:
\beq\label{coplanarity}
b = \sqrt{b^2_1 + b^2_2}, \qquad \text{where $b_i = \hat{\epsilon}_i\cdot\hat{\mu}_\Sun$,}
\eeq
and $\hat{\mu}_\Sun$ is the unit vector pointing in the direction of the DM wind in the heliocentric frame:
\beq\label{DM_wind_sb_sun}
\vec{\mu}_\Sun = \vec{\mu} - \vec{v}_{\Sun}.
\eeq
The definition in Eq.~\eqref{coplanarity} is equivalent to $\sin \lambda$ with $\lambda$ the angle between the normal of the Earth orbital plane and $\hat{\mu}_\Sun$.

\section{Phase inversion in scattering rate}
\label{appC}

As discussed in Sec.~\ref{subsubsec:modulation} and illustrated in Fig.~\ref{fig:gvmin_tdep}, the modulation of the plateau and the tail of $g(\vmin, t)$ have opposite phases, minimized and maximized at the characteristic time $t_c$, respectively. Since $E$ can be mapped onto $\vmin$, one would expect this same behavior to be observed in $\kappa(E,t)$ and consequently the scattering rate.

\begin{figure*}[tbh]
    \centering
        \includegraphics[width=0.48\textwidth]{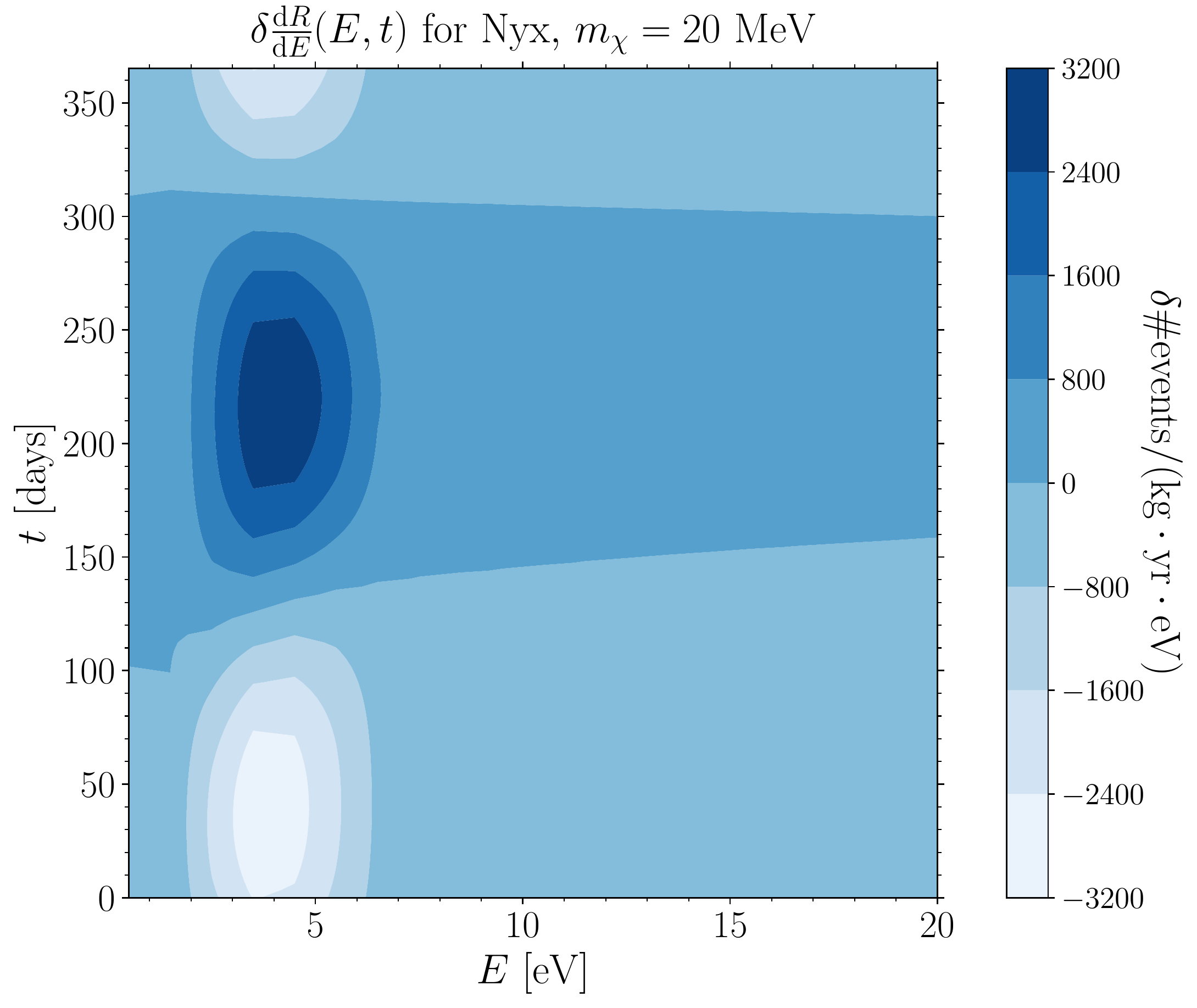}
        \includegraphics[width=0.48\textwidth]{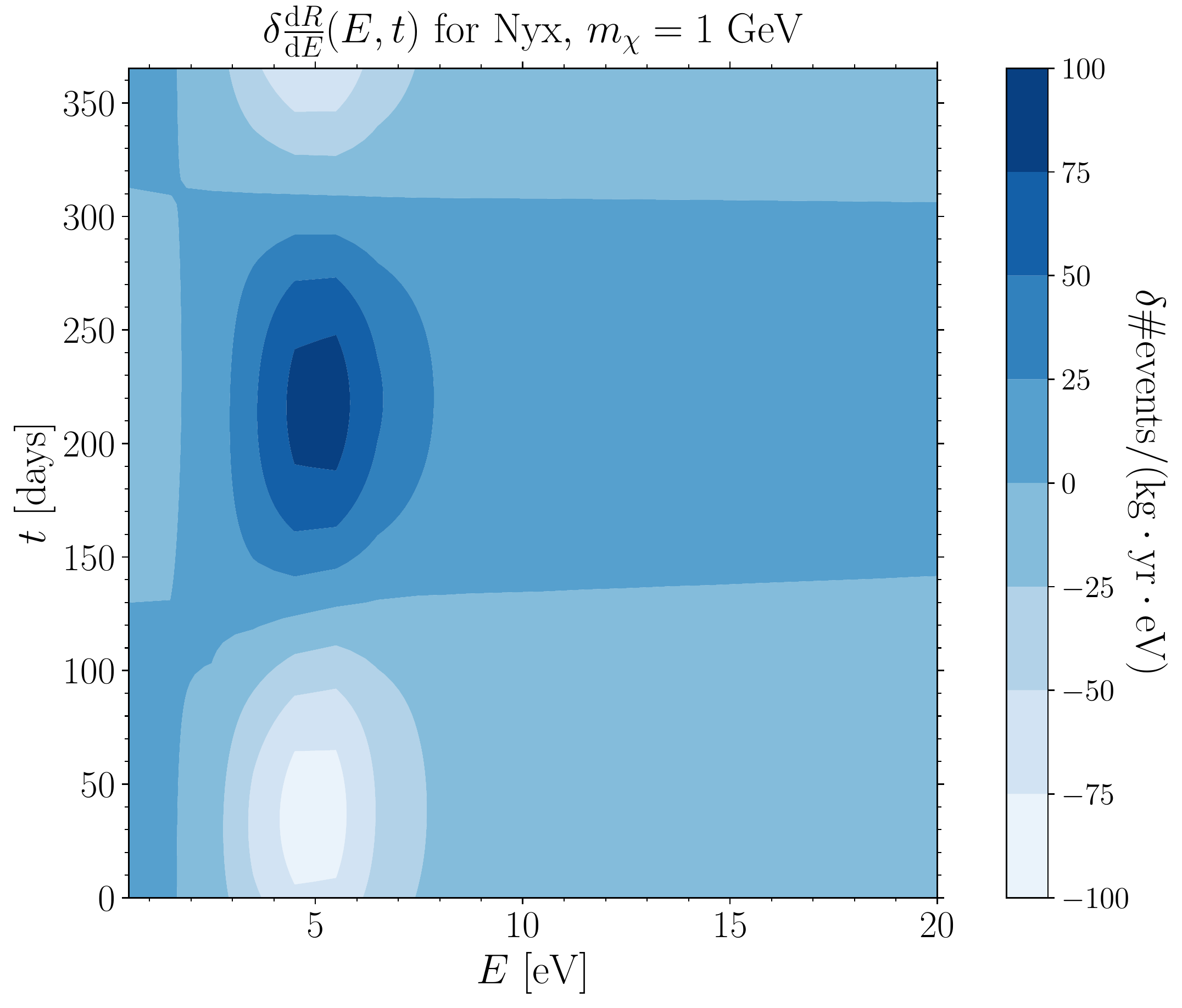}
    \caption{{\it Left:} Spectrum modulation $\delta \frac{\dd R}{\dd E} (E, t)$ for 100\% of the local DM coming from Nyx, for $m_\chi=20~\MeV$ ({\it left}) and $m_\chi=1~\GeV$ ({\it right}). Note that in the low mass case, there is only one modulation phase while in the high mass case, there are two opposite phases at low and high $E$'s.}
    \label{fig:inverse_phase}
\end{figure*}

However, as shown in the right panel of Fig.~\ref{fig:f2_gvmin}, whether a given $E$ gives a $\vmin$ in the plateau or the tail of $g(\vmin, t)$ depends on the DM mass $m_\chi$: lower masses yield large $\vmin$'s which are entirely within the tail, particularly if $g(\vmin, t)$ is ``narrow'', as in the case for DM components with low $\vmp$'s, such as Nyx. As a consequence, the plateau modulation will only be present in the scattering spectra for DM with large masses. For small $m_\chi$, only the tail modulation will be observed in the spectra. Fig.~\ref{fig:inverse_phase} illustrates this for the case of Nyx, for $m_\chi=20~\MeV$ and $m_\chi=1~\GeV$.

\section{Additional results}
\label{appD}

\begin{figure*}[tbh]
    \centering
         \includegraphics[width=0.48\textwidth]{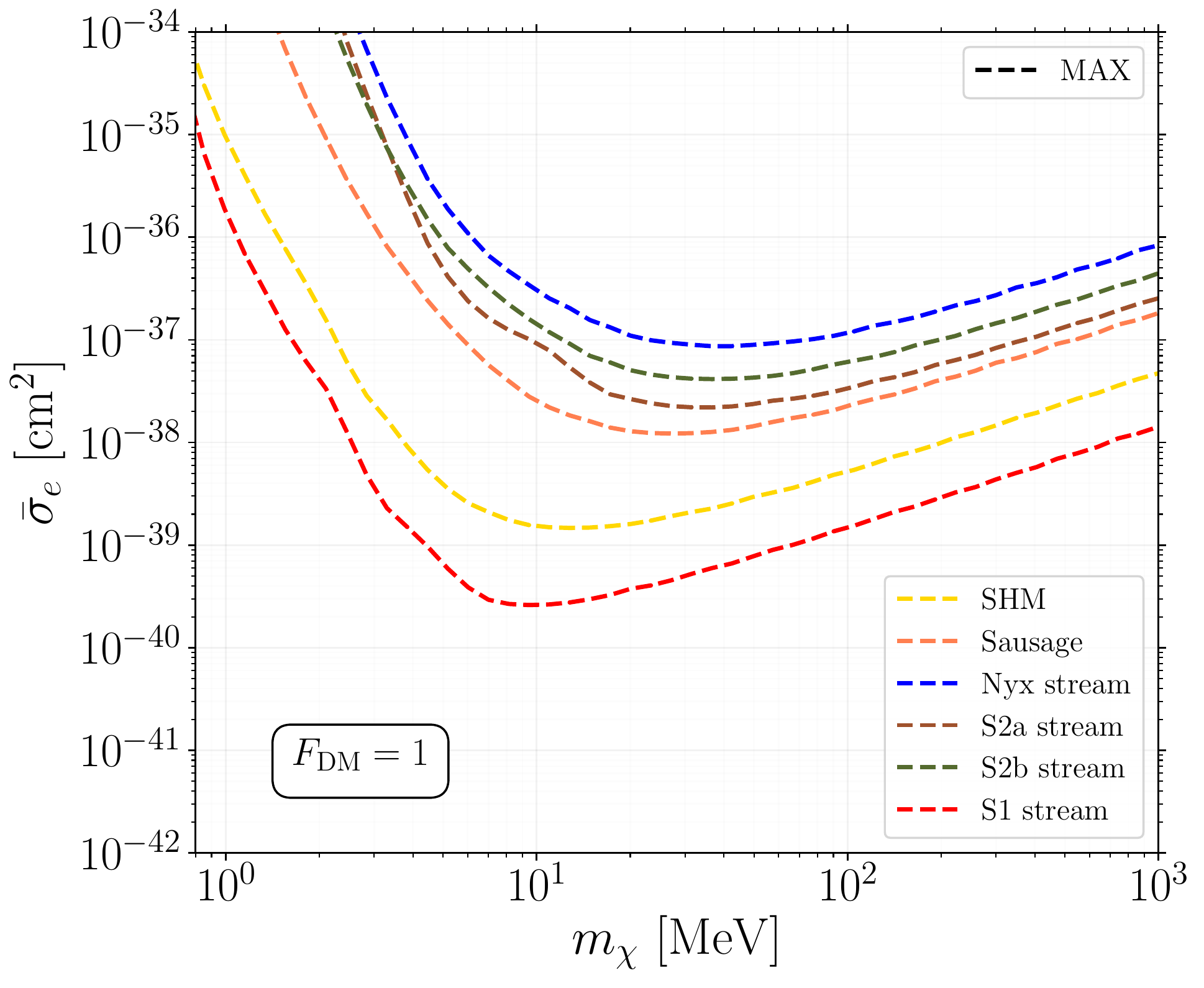}
         \includegraphics[width=0.48\textwidth]{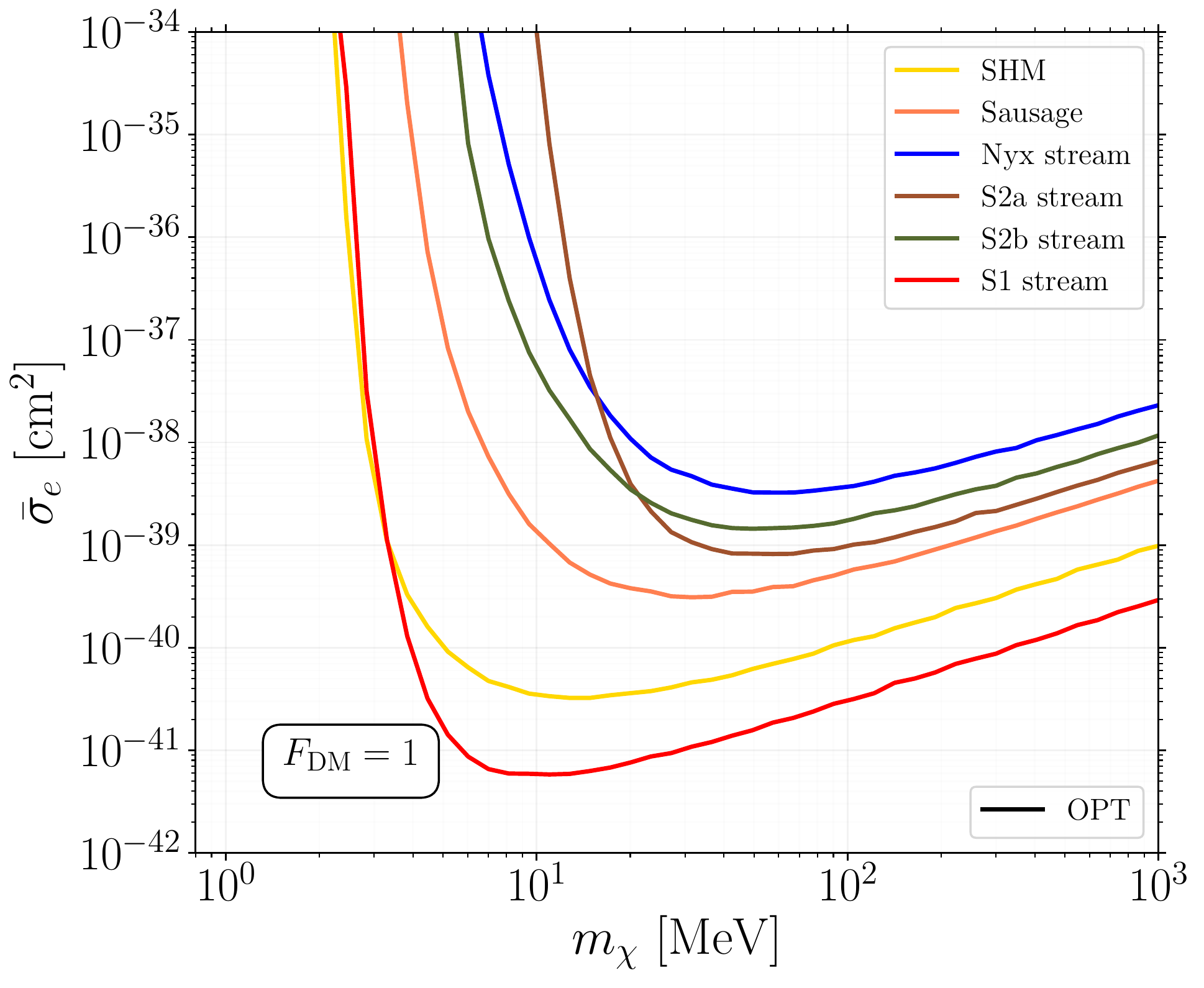}
    \caption{5$\sigma$ discovery reaches for a non-modulating DM signal using MAX ({\it left}) and OPT ({\it right}) background model for various DM substructure components discussed in Sec.~\ref{sec:discreaches}.}
    \label{fig:dr_compare_app}
\end{figure*}

In this appendix, we include additional results that further illustrate the effects DM components could have on a SENSEI-like direct detection experiment. Fig.~\ref{fig:dr_compare_app} shows the $5\sigma$ discovery reaches for DM with $n=0$ ($F_\DM = 1$) scattering off electrons in a silicon-based experiment, for two background models, MAX and OPT, described in Table~\ref{tab:dd_summary}. For each contour, we assume that 100\% of local DM particles are drawn from the corresponding astrophysical component: SHM, {\it Gaia} Sausage, Nyx, S1, S2a, or S2b. Note that since these are discovery reaches for non-modulating DM signals, the characteristic time $t_c$ and the coplanarity $b$ do not play a role.

In both background models, the high DM mass behavior of the reaches can be understood in terms of the $\vmp$ values of the distributions: the components with larger $\vmp$'s relative to SHM present a stronger discovery reach (see Table~\ref{table:vmp_t0}). Indeed, since at high DM masses $\vmin$ is very small, the entire width of their respective $\overline{g}(\vmin)$ is available for scattering, and as shown in Fig.~\ref{fig:f2_gvmin}, they receive a larger contribution from the crystal form factor at low energies. However, for those components with lower $\vmp$'s, their higher $\overline{g}(\vmin)$ relative to SHM is not enough to overcome the smaller form factor contribution, resulting in a weaker discovery reach. This effect is compounded by a low signal-to-noise ratio (no signal) in $Q= 1,2$ bins, where the spectra for components with lower $\vmp$'s peak, for the MAX (OPT) background model. Finally, we note that the relative strength of the OPT case compared to the MAX one is determined by the fact that the former has a lower constant background rate than the latter.

For low DM masses, $\vmin$ becomes larger, thereby making only the tail of $\overline{g}(\vmin)$ available for scattering. As a consequence, those distributions with both larger $\vmp$ (which allows for a wider of $\overline{g}(\vmin)$) and velocity dispersion $\sigma_v$ have a stronger reach. This is more evident in the right panel of Fig.~\ref{fig:dr_compare_app} corresponding to the OPT background model, since the ionization threshold $Q_{\rm th}=3$ means that the narrow distributions with low $\vmp$ have to rely on their velocity dispersions $\sigma_v$ to produce any events.

\bibliography{references}

\end{document}